\documentclass[aps,prd,twocolumn,nofootinbib,preprintnumbers,showkeys,superscriptaddress]{revtex4}

\usepackage{newtxtext}
\usepackage{newtxmath}
\usepackage{mathrsfs}

\usepackage{graphicx}
\usepackage{bm}
\usepackage{amsmath}
\usepackage{float}
\usepackage{multirow}
\usepackage{slashed}
\usepackage{xcolor}
\usepackage{mathtools,braket}
\usepackage{color,physics}

\usepackage[mathlines]{lineno}

\def\orcid#1{\kern .08em\href{https://orcid.org/#1}{\includegraphics[keepaspectratio,width=0.7em]{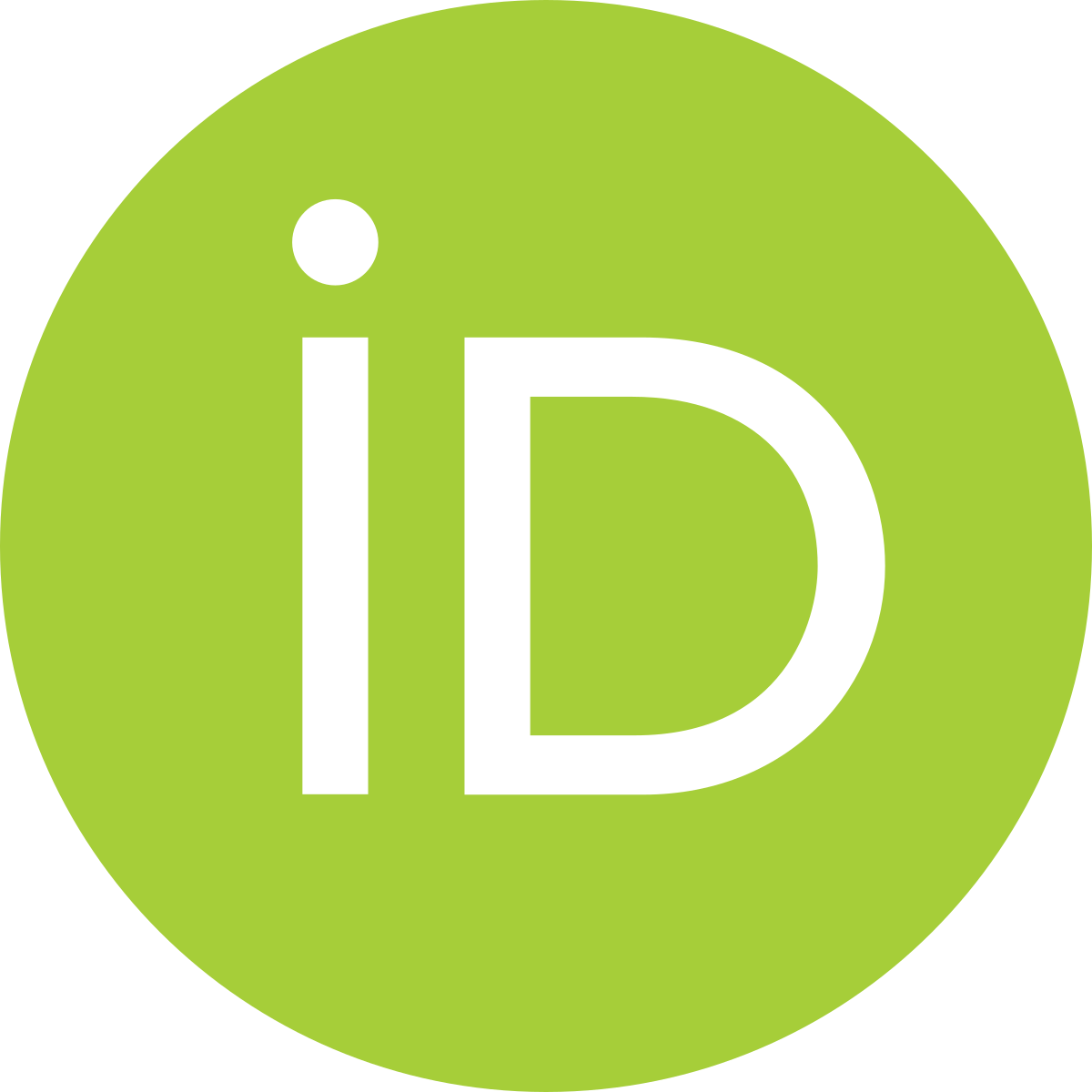}}}

\usepackage[colorlinks=true, pdfstartview=FitV, bookmarks=true, bookmarksnumbered=true, breaklinks]{hyperref}
\hypersetup{linkcolor=blue, citecolor=blue, urlcolor=blue}

\usepackage[T1]{fontenc}
\usepackage{amsmath}

\usepackage{color}
\definecolor{blue}{rgb}{0.0, 0.0, 1.0}
\definecolor{red}{rgb}{1.0, 0.0, 0.0}
\definecolor{royalblue}{rgb}{0.0, 0.14, 0.4}
\hypersetup{linkcolor= blue, citecolor=blue, urlcolor=blue}

\def\orcid#1{\kern .08em\href{https://orcid.org/#1}{\includegraphics[keepaspectratio,width=0.7em]{ORCID_iD.png}}}

\begin{document}

\title{\boldmath
Mixing effects on $1S$ and $2S$ state heavy mesons in the light-front quark model}

\author{Ahmad Jafar Arifi\orcid{0000-0002-9530-8993}}
\email{ahmad.jafar.arifi@apctp.org}
\affiliation{Asia Pacific Center for Theoretical Physics, Pohang, Gyeongbuk 37673, Korea}

\author{Ho-Meoyng Choi\orcid{0000-0003-1604-7279}}
\email{homyoung@knu.ac.kr}
\affiliation{Department of Physics Education, Teachers College, Kyungpook National University, Daegu 41566, Korea}

\author{Chueng-Ryong Ji\orcid{0000-0002-3024-5186}}
\email{ji@ncsu.edu}
\affiliation{Department of Physics, North Carolina State University, Raleigh, NC 27695-8202, USA}

\author{Yongseok Oh\orcid{0000-0001-9822-8975}\ \  }
\email{yohphy@knu.ac.kr}
\affiliation{Department of Physics, Kyungpook National University, Daegu 41566, Korea}
\affiliation{Asia Pacific Center for Theoretical Physics, Pohang, Gyeongbuk 37673, Korea}

\date{\today}

\begin{abstract}
The mass spectra and wave functions of both $1S$ and $2S$ state heavy pseudoscalar ($P$) and vector ($V$) mesons 
are analyzed within the light-front quark model. 
Important empirical constraints employed in our analysis of the mass spectra and wave functions are the experimental mass-gap relation, 
$\Delta M_P > \Delta M_V$, where $\Delta M_{P(V)}=M^{2S}_{P(V)}-M^{1S}_{P(V)}$
and the hierarchy of the decay constants, $f_{1S}>f_{2S}$, between $1S$ and 
$2S$ meson states. We maintain the orthogonality of the trial wave functions of 
the $1S$ and $2S$ states in our variational calculation of the Hamiltonian with the 
Coulomb plus confining potentials and treat the hyperfine interaction perturbatively
for the heavy-heavy and heavy-light $P$ and $V$ mesons due to the nature of the heavy quark symmetry.
Realizing that the empirical constraints cannot be satisfied without mixing of the $1S$ and $2S$ states, we find the lower bound 
of the mixing angle $\theta$ between $1S$ and $2S$ states as $\theta_c = {\rm cot}^{-1}(2\sqrt{6})/2\simeq 6^{\circ}$ 
and obtain the optimum value of the mixing angle around $12^\circ$ to cover both the charm and bottom 
flavors of the heavy quark. 
The mixing effects are found to be more significant to the $2S$ state mesons than to the $1S$ state mesons.
The properties of $1S$ and $2S$ state mesons including the mass spectra, decay constants, twist-2 distribution amplitudes, 
and electromagnetic form factors are computed.
Our results are found to be in a good agreement with the available data and lattice simulations.
In particular, the $2S$ state pseudoscalar $D_s$ meson is predicted to have a mass of $2600$~MeV, which is very close 
to the mass of the newly discovered $D_{s0}(2590)^+$ meson by the LHCb Collaboration.
This supports the interpretation of the observed state as a radial excitation of the $D^+_s$ meson.
\end{abstract}

\maketitle

\section{Introduction}

Quantum chromodynamics (QCD) is a unique theory of strong interactions with its non-perturbative nature in the low-energy regime 
and its asymptotically free nature in the high-energy regime. 
Building the effective degrees of freedom that describe the strongly interacting system in the low-energy regime is one of the crucial 
issues in understanding the link between the first-principle QCD and the constituent quark model (CQM) that has proven to provide 
successful and intuitive descriptions of hadrons. 
In particular, the light-front dynamics (LFD) is found to provide an effective way to handle the relativistic effects thanks to its distinguished 
features of the rational energy-momentum dispersion relation. 
It carries the maximum number (seven) of the kinetic (or interaction-independent) generators rendering the less effort in dynamics to get 
the QCD solutions that reflect the full Poincar\'e symmetries~\cite{Dirac49,BPP97}. 
Effectively, the light-front quark model (LFQM) based on the LFD turns out to be one of the most successful hadronic models in describing 
various properties of hadrons.

While the LFQM analyses have been quite successful in describing the properties of ground state 
mesons~\cite{Jaus90,Jaus99,CCHZ97,Hwang01,KLWL10,CLLSY18,AJ99,CJ97,CJ99a,CJ09,Choi07,CJLR15,DDJC19}, 
the structures and properties of the excited hadron states are yet to be understood in LFQM more extensively as their nature is still 
veiled and not well explored compared to the ground states. 
One of the most challenging problems in the quest of the excited states is to clarify whether the observed state belongs to the standard 
quark-antiquark excitation or an exotic state.
For instance, the newly observed $D_{s0}(2590)^+$ meson with a mass of $2591 \pm 6 \pm 7$~MeV by the LHCb Collaboration~\cite{LHCb-20d} 
has been proposed as a radial excitation of the $D^+_s$ meson. 
However, the observed mass is quite smaller than the available CQM predictions. 
For example, the relativized quark model of Ref.~\cite{GI85} predicts $2680$~MeV and the relativistic quark model based on the 
quasipotential approach predicts $2688$~MeV~\cite{EFG02b,EFG09b}. 
In recent works, therefore, some nonstandard quark-antiquark behaviors attributed to this resonance have been discussed~\cite{OSEF21,XLG21}.

In particular, the radially excited states of hadrons are important in understanding the strong interactions as they give information complementary 
to the orbitally excited states.
They have been observed in light and heavy quark sectors of hadrons although some of them are yet to be confirmed according to the 
Particle Data Group (PDG)~\cite{PDG20}.
The most well-known example in baryon spectrum would be the Roper resonance~\cite{Roper64}.
Such states with multistrangeness were recently discussed in Ref.~\cite{ASHO22}, and
the excited states in meson spectrum were discussed and summarized, for example, in Ref.~\cite{KZ07}. 
Most notable empirical hierarchy appears in the radially excited $2S$ state as well as the ground $1S$ state of heavy pseudoscalar ($P$) 
and vector ($V$) mesons. 
Namely, the two important constraints that we notice from the empirical hierarchy are (i) the experimental mass gap relation 
$\Delta M_P > \Delta M_V$, where $\Delta M_{P(V)}=M^{2S}_{P(V)}-M^{1S}_{P(V)}$, 
and (ii) the hierarchy of the decay constants $f_{1S}> f_{2S}$. 
While these constraints on the mass spectra and the decay constants apply both for the light and heavy meson sectors, the difference 
between $P$ and $V$ becomes much larger in the light meson sectors.
The reason for the smaller difference between $P$ and $V$ in the heavy-light system may well be attributed to the 
heavy quark symmetry~\cite{VS87,VS88,IW89,IW90} as well as the perturbative nature of the hyperfine interaction in heavy quark systems. 
This may be contrasted with the chiral symmetry reflected in the light meson sectors, which deserves a separate analysis and discussion. 
Due to these significant differences in the underlying symmetry between the light and heavy meson sectors, we apply the 
two constraints here only for the heavy meson sectors and discuss the heavy-heavy and heavy-light $P$ and $V$ mesons in the present work. 
Effectively, we analyze the mass spectra and wave functions of the radially excited $2S$ state and the ground $1S$ 
state of heavy $P$ and $V$ meson sectors within the framework of the LFQM and discuss various properties of these mesons.

In the previous LFQM analyses of the mass spectra and decay constants of the $1S$ state mesons performed by two of us with 
the QCD-motivated effective Hamiltonian~\cite{CJ97,CJ99a,CJ09,Choi07,CJLR15,DDJC19}, the trial wave functions were chosen as 
either the pure harmonic oscillator (HO) wave function $\phi_{1S}$~\cite{CJ97,CJ99a,CJ09,Choi07} or an expansion in the HO basis functions, i.e., 
$\Phi = \sum^{n_{\rm max}}_{n=1}c_n \phi_{nS}$ with $n_{\rm max}=2$~\cite{CJLR15} or 3~\cite{DDJC19}. 
Through the analyses of the $1S$ state mesons, it was shown that the physical observables are not much sensitive to the number of HO bases 
used in the trial wave functions, $\Phi = \sum^{n_{\rm max}}_{n=1}c_n \phi_{nS}$, once the optimum values of the model parameters are fitted.
For the combined analysis of $(1S, 2S)$ state heavy mesons in the present work, we take into account the two important constraints in 
obtaining the optimum values of our model parameters.
We note that $f_{nS}$ tends to be smaller as $n$ gets larger since the decay constant of a hadron is proportional to its radial wave function 
at the origin, $\psi(r=0)$.
The available experimental data have also confirmed this tendency.
In the literature, however, some difficulties in the combined analysis of the ground and radially excited states have been observed. 
For instance, in the LFQM analysis of ($1S,2S$) state heavy $\Upsilon(b\bar{b})$ systems, it was found that using the HO wave functions 
$\phi_{nS}(n=1,2)$ leads to the reverse order problem of $f_{1S} < f_{2S}$~\cite{KLWL10}.
In order to resolve this problem and to obtain the correct hierarchy of $f_{1S} > f_{2S}$ for the heavy quarkonium system using 
two pure $(\phi_{1S}, \phi_{2S})$ wave functions, the authors of Refs.~\cite{Hwang08,PM12} had to choose different HO model 
parameters for different $nS$ states, which breaks the orthogonality condition between the two wave functions $\phi_{1S}$ 
and $\phi_{2S}$.
A similar problem, namely, the breakdown of orthogonality condition between the resultant $1S$ and $2S$ state light-front (LF) 
wave functions, appears in the analysis of heavy quarkonium system performed with the basis light-front quantization (BLFQ) 
approach in a holographic basis~\cite{LLCLV21}.

The main purpose of the present work is thus to extend the previous LFQM analyses including both $1S$ and $2S$ state $P$ and $V$ 
heavy meson sectors to remedy the difficulties in the combined analysis using the trial wave functions for $1S$ and $2S$ states as 
mixtures of the two HO wave functions $\phi_{1S}$ and $\phi_{2S}$. 
In particular, our trial wave functions for $1S$ and $2S$ states satisfy naturally the orthogonality condition. 
One of the key findings in this work is 
the criterion of the mixing angle between $\phi_{1S}$ and $\phi_{2S}$
for reproducing the correct order of the mass gap (i.e., $\Delta M_P > \Delta M_V$) and decay constants (i.e., $f_{1S} > f_{2S}$) 
between $1S$ and $2S$ states.
Various properties of heavy ($1S, 2S$) state mesons such as mass spectra, decay constants, distribution amplitudes (DAs), and 
electromagnetic form factors are also scrutinized.
Moreover, we obtain the mass of the radially excited $D_{s}(2S)$ state as $M \approx 2600$~MeV, which leaves the possibility that 
the $D_{s0}(2590)^+$ observed recently by the LHCb Collaboration~\cite{LHCb-20d} can be interpreted as the standard 
quark-antiquark radial excitation.

This paper is organized as follows.
In Sec.~\ref{sec:model}, we briefly introduce the effective Hamiltonian and the trial wave functions adopted in the present
approach.
We also describe how to determine the model parameters via the variational analysis.
Subsequently, we describe the mass spectra of the $1S$ and $2S$ state heavy mesons and the role of the mixing is 
discussed as well.
In Sec.~\ref{sec:applications}, we summarize various properties of heavy mesons including decay constants, DAs, 
and electromagnetic form factors obtained in our LFQM formalism.
Section~\ref{sec:results} presents our numerical results for those quantities of $(1S, 2S)$ state heavy pseudoscalar and 
vector mesons.
Finally, we summarize and conclude in Sec.~\ref{sec:summary}.

\section{Model description} \label{sec:model}

The key idea in our LFQM~\cite{CJ97,CJ99a,CJ09,CJLR15,Choi07} for the $1S$ state mesons is to treat the radial 
wave function as a trial function for the variational principle to the QCD-motivated effective Hamiltonian saturating 
the Fock state expansion by the constituent quark and antiquark.
In this section, we briefly summarize our LFQM and discuss some distinguished features for the trial wave functions 
and their mixing angles that emerge from the inclusion of the radially excited $2S$ state in addition to the $1S$ state.

\subsection{Effective Hamiltonian}

The meson system at rest is described as an interacting bound system of effectively dressed valence quark  
and antiquark satisfying the eigenvalue equation of the QCD-motivated effective Hamiltonian,
\begin{eqnarray}\label{eq:1}
	H_{q\bar{q}} \ket{\Psi_{q\bar{q}}} = M_{q\bar{q}} \ket{\Psi_{q\bar{q}}},
\end{eqnarray}
where $M_{q\bar{q}}$ and $\Psi_{q\bar{q}}$ are the mass eigenvalue and eigenfunction of the $q\bar{q}$ meson state, respectively. 
We take the Hamiltonian $H_{q\bar{q}}$ in the quark-antiquark center of mass frame as 
\begin{equation}\label{eq:2}
H_{q\bar{q}} = H_0 + V_{q\bar{q}} = \sqrt{m_q^2 + {\bf k}^2}  +  \sqrt{m_{\bar{q}}^2 + {\bf k}^2} + V_{q\bar{q}}, 
\end{equation}
where  $H_0$ is the kinetic energy part of the quark and antiquark 
with three-momentum ${\bf k}=({\bf k}_\perp, k_z)$. 
The effective potential $V_{q\bar{q}}$ is given by~\cite{CJ97,CJ99a,CJ09,Choi07}
\begin{eqnarray}
V_{q\bar{q}} &=&  V_{\rm Conf} + V_{\rm Coul} + V_{\rm Hyp},
\end{eqnarray}
where $V_{\rm Conf}$ is the linear confining potential,
\begin{equation}
V_{\rm Conf} = a + br
\end{equation}
with $a$ and $b$ being parameters to be determined later.
The Coulomb potential and hyperfine interaction potential stemming from the effective one-gluon exchanges 
for the $S$-wave mesons are written as
\begin{equation} 
V_{\rm Coul} = -\frac{4\alpha_s}{3r}, \qquad
V_{\rm Hyp} = \frac{2}{3}\frac{\mathbf{S}_q \cdot \mathbf{S}_{\bar{q}}}{m_q m_{\bar{q}}} \bm{\nabla}^2 V_{\rm Coul}.
\end{equation}
We take the strong coupling constant $\alpha_s$ as a parameter and $\Braket{ \mathbf{S}_q \cdot \mathbf{S}_{\bar{q}} }$ 
is $1/4$ and $-3/4$ for vector and pseudoscalar mesons, respectively.
As we consider the heavy meson sector in this work, we handle $V_{\rm Hyp}$ perturbatively employing the 
contact hyperfine interaction, i.e.,
$\bm{\nabla}^2 V_{\rm Coul} = (16\pi \alpha_s/3) \delta^3({\bf r})$, which is a fairly good approximation for the analysis of heavy meson mass spectroscopy.

The LF wave function is represented by the Lorentz invariant internal variables $x_i = p^+_i /P^+$, 
${\bf k}_{\perp i} = {\bf p}_{\perp i} - x_i {\bf P}_{\perp}$, and helicity $\lambda_i$, where 
$P^\mu = (P^+,P^-,\mathbf{P}_\perp)$ is the four-momentum of the meson and $p^\mu_i$ is the four-momentum of 
the $i$th ($i=1,2$) constituent quark, which leads to the constraints $\sum^2_{i=1} x_i = 1$ and 
$\sum^2_{i=1}{\bf k}_{\perp i} = 0$.
We assign $i=1$ to the quark and $i=2$ to the antiquark, and define $x \equiv x_1$ with ${\bf k}_\perp \equiv {\bf k}_{\perp 1}$.
Then the three-momentum ${\bf k} = (k_z, {\bf k}_\perp)$ can be written as ${\bf k} = (x, {\bf k}_\perp)$
via the relation,
\begin{equation}
k_z = \left( x- \frac12 \right) M_0 + \frac{m^2_{\bar q}-m^2_q}{2M_0} , 
\end{equation}
where
\begin{eqnarray}\label{eq:4}
	M_0^2 = \frac{\mathbf{k}_{\bot}^2 + m_q^2}{x}  + \frac{\mathbf{k}_{\bot}^2 + m_{\bar{q}}^2}{1-x}
\end{eqnarray}
is the boost-invariant meson mass squared. 
Therefore, the variable transformation $\{k_z, \mathbf{k}_\bot \} \to \{x, \mathbf{k}_\bot \}$ accompanies
the Jacobian factor, 
\begin{equation}
\frac{\partial k_z}{\partial x} = \frac{M_0}{4x(1-x)} \left[ 1 - \frac{ (m_q^2 - m_{\bar{q}}^2)^2}{M_0^4} \right],
\end{equation}
which we take into account for the normalization of the radial part of the wave function.

The LF wave function, $\Psi_{q{\bar q}}=\Psi^{JJ_z}_{nS}$ of the $nS$ state pseudoscalar and vector mesons, 
in momentum space is then given by 
\begin{eqnarray}\label{eq:5}
		\Psi^{JJ_z}_{nS}(x, \mathbf{k}_{\bot},\lambda_i) = \Phi_{nS}(x, \mathbf{k}_\bot)
		\  \mathcal{R}^{JJ_z}_{\lambda_q\lambda_{\bar{q}}}(x, \mathbf{k}_\bot),
\end{eqnarray}
where $\Phi_{nS}(x, \mathbf{k}_\bot)$ is the radial wave function and $\mathcal{R}^{JJ_z}_{\lambda_q\lambda_{\bar{q}}}$ 
is the spin-orbit wave function that is obtained by the interaction-independent Melosh transformation from the ordinary
spin-orbit wave function assigned by the quantum number $J^{PC}$.
The covariant forms of $\mathcal{R}^{JJ_z}_{\lambda_q\lambda_{\bar{q}}}$ for pseudoscalar and vector mesons are 
given by~\cite{Jaus90}
\begin{eqnarray}\label{eq:6}
	\mathcal{R}^{00}_{\lambda_q\lambda_{\bar{q}}} &=&  -\frac{1}{\sqrt{2} \tilde{M}_0} 
	\bar{u}_{\lambda_q}^{}(p_q) \gamma_5  v_{\lambda_{\bar{q}}}^{}(p_{\bar{q}}), 
	\nonumber\\
	\mathcal{R}^{1 J_z}_{\lambda_q\lambda_{\bar{q}}} &=& -\frac{1}{\sqrt{2} \tilde{M}_0} 
	\bar{u}_{\lambda_q}^{}(p_q) 
	\left[ \slashed{\epsilon}(J_z) - \frac{\epsilon \cdot (p_q-p_{\bar{q}})}{M_0 + m_q + m_{\bar{q}}} \right] 
	v_{\lambda_{\bar{q}}}^{}(p_{\bar{q}}),
	\nonumber \\
\end{eqnarray}
where $\tilde{M}_0 \equiv \sqrt{M_0^2 - (m_q -m_{\bar{q}})^2}$. 
The polarization vectors $\epsilon^\mu(J_z)=(\epsilon^+, \epsilon^-,\bm{\epsilon}_{\perp})$ of the vector meson 
are given by~\cite{Jaus90}
\begin{eqnarray}\label{eq:7}
\epsilon^\mu(\pm 1) &=& \left( 0, \frac{2}{P^+} \bm{\epsilon}_\perp(\pm) \cdot {\bf P}_\perp, \bm{\epsilon}_\perp(\pm)\right),
\nonumber\\
\epsilon^\mu(0) &=& \frac{1}{M_0}\left(P^+, \frac{-M^2_0 + {\bf P}^2_\perp}{P^+}, {\bf P}_\perp\right),
\end{eqnarray}
where
\begin{equation}
\bm{\epsilon}_\perp(\pm 1) = \mp \frac{1}{\sqrt{2}} \left( 1, \pm i \right),
\end{equation}
so that the spin-orbit wave functions $\mathcal{R}^{JJ_z}_{\lambda_q\lambda_{\bar{q}}}$ satisfy the 
unitary condition automatically, i.e.,
$\Braket{ \mathcal{R}^{JJ_z}_{\lambda_q\lambda_{\bar{q}}} | \mathcal{R}^{JJ_z}_{\lambda_q\lambda_{\bar{q}}} } = 1$.

For the $1S$ and $2S$ state radial wave functions $\Phi_{ns}$ of Eq.~(\ref{eq:5}), 
we allow the mixing between the two lowest order HO wave functions ($\phi_{1S},\phi_{2S}$) by writing
\begin{eqnarray}\label{eq:8}
	\begin{pmatrix} \Phi_{1S} \\ \ \Phi_{2S}  \end{pmatrix} = 
	\begin{pmatrix} \cos\theta & \sin \theta  \\ -\sin \theta & \cos \theta \end{pmatrix} 
	\begin{pmatrix} {\phi}_{1S}  \\ {\phi}_{2S} \end{pmatrix},
\end{eqnarray}
where 
\begin{eqnarray}\label{eq:9}
	\phi_{1S} (x, \mathbf{k}_\bot) &=& \frac{4\pi^{3/4}}{ \beta^{3/2}} 
	\sqrt{\frac{\partial k_z}{\partial x}} e^{-{\bf k}^2/ 2\beta^2},
	\nonumber\\
	\phi_{2S} (x, \mathbf{k}_\bot) &=& \frac{4\pi^{3/4}}{ \sqrt{6}\beta^{7/2}} 
	\left( 2 {\bf k}^2 - 3\beta^2 \right) \sqrt{\frac{\partial k_z}{\partial x}} e^{-{\bf k}^2/ 2\beta^2},
\end{eqnarray}
and $\beta$ is the parameter which is inversely proportional to the range of the wave function
and can be used as the variational parameter in our mass spectroscopic analysis. 
It should be noted
that the wave functions $\phi_{nS}$ include the Jacobian factor $\partial k_z/\partial x$ so that
the HO bases $\phi_{nS}$ satisfy the following normalization:
\begin{eqnarray}\label{eq:10}
 \int_0^1  dx \int \frac{d^2 \mathbf{k}_\bot}{2(2\pi)^3}  \abs{ \phi_{nS}(x, \mathbf{k}_\bot) }^2 =1.
\end{eqnarray}
From the orthonormality of $\Phi_{nS}(n=1,2)$ defined in Eq.~(\ref{eq:8}) and the unitarity of 
$\mathcal{R}^{JJ_z}_{\lambda_q\lambda_{\bar{q}}}$, one can easily see that $\Phi_{nS}$ and $\Psi_{nS}^{J J_z}$ 
of Eq.~(\ref{eq:5}) satisfy the same normalization as $\phi_{nS}$.
We denote $(\Phi_{1S}, \Phi_{2S})$ for $\theta\neq 0$ and $(\Phi_{1S}, \Phi_{2S}) = (\phi_{1S}, \phi_{2S})$ 
for $\theta=0$ as ``mixed'' and ``pure'' ($1S$, $2S$) states, respectively.
As we shall discuss below, the mixing scheme turns out to be crucial to reproduce the experimental data for 
both masses and decay constants of heavy mesons.

\subsection{Variational method to effective Hamiltonian}

The present LFQM for the combined analysis of the $1S$ and $2S$ state heavy mesons has several parameters, 
namely, the constituent quark masses $(m_q, m_s, m_c, m_b)$ with $m_q$ being the light $u$ and $d$ quark mass, 
the potential parameters $(a, b,\alpha_s)$, the HO parameter $\beta$ for each $(q\bar{q})$ content, and the mixing 
angle $\theta$. 
We first determine the values of these parameters by reproducing the mass spectra based on the variational principle.
Then we compute other observables of heavy mesons such as decay constants, distribution amplitudes (DAs), and 
electromagnetic form factors.

Here we follow the procedure adopted in Refs.~\cite{CJ97,CJ99a,Choi07,CJ09}, namely, we consider the central 
potential $V_0=V_{\rm Conf}+V_{\rm Coul}$ as well as the kinetic energy $H_0$ in variational calculation via
\begin{eqnarray}\label{eq:11}
	\frac{\partial \bra{\Psi_{q\bar q}} \left( H_0 + V_0\right) \ket{\Psi_{q\bar q}} }{\partial \beta} = 0.
\end{eqnarray}
Then the remaining $\Braket{\Psi_{q\bar q} | V_{\rm hyp} | \Psi_{q\bar q}}$ is treated as a perturbation so that we 
have $\beta$ values common for both pseudoscalar and vector mesons of the same $(q\bar q)$ content. 
This constrains the model parameters.
Since the spin-orbit wave function satisfies the exact unitarity, we have the mass eigenvalue of the meson as 
$M_{q\bar{q}}=\Braket{\Psi_{q\bar q}| H_{q\bar q}|\Psi_{q\bar q} }= \Braket{ \Phi_{nS}| H_{q\bar q}|\Phi_{nS} }$.
The analytic forms of the mass eigenvalues $(M_{q\bar{q}}^{1S}, M_{q\bar{q}}^{2S}$) for the mixed ($1S$, $2S$) 
state mesons are then obtained as
\begin{widetext}
\begin{eqnarray}\label{eq:12}
	M_{q\bar{q}}^{1S} &=& \frac{\beta}{ \sqrt{\pi}} \sum_{i=q,\bar{q}} \biggl\{ z_i e^{z_i/2} \biggl[  \frac{1}{3} c_2^2 (3-z_i) z_i~K_2\left(\frac{z_i}{2}\right)   
	+\frac{1}{6}  \left(9 - 3c_1^2 + 2c_2^2z_i^2 -6\sqrt{6}c_1 c_2\right)K_1\left(\frac{z_i}{2}\right)  \biggr] 
	\nonumber\\ & & \mbox{} \qquad \qquad
	+ \sqrt{\pi} \left(\sqrt{6}c_1 c_2 - 3c_2^2\right) U\left(-1/2,-2,z_i\right)\biggr\}	 
	\nonumber \\ && \mbox{} 
	+ a 
	+ \frac{b}{\beta\sqrt{\pi}} \left(3- c_1^2 -2 \sqrt{\frac{2}{3}}c_1 c_2 \right)
	- \frac{4\alpha_s \beta}{9\sqrt{\pi}} \left( 5+c_1^2 + 6\sqrt{\frac{2}{3}}c_1 c_2 \right)  
	+  \frac{16  \alpha_s \beta^3  \Braket{ \mathbf{S}_q\cdot \mathbf{S}_{\bar{q}} } }{9m_q m_{\bar{q}} \sqrt{\pi}} 
	(3-c_1^2 + 2\sqrt{6}c_1 c_2),
	\nonumber\\
	M_{q\bar{q}}^{2S} &=& M_{q\bar{q}}^{1S} (c_1\to -c_2, c_2\to c_1),
\end{eqnarray}
\end{widetext}
where $(c_1,  c_2)=(\cos\theta, \sin\theta)$, $z_i = m_i^2/\beta^2$, $K_n(x)$ is the modified Bessel function of the second kind of order $n$,
and $U(a,b,z)$ is the Tricomi's (confluent hypergeometric) function.
The mass eigenvalues for the pure ($1S$, $2S$) states can be read by setting $\theta = 0$, i.e., ($c_1=1$, $c_2=0$) in Eq.~(\ref{eq:12}).

In order to explore the mixing effects and to determine the optimal value of the mixing angle $\theta$, we utilize the empirical constraint 
on the mass gap $\Delta M_{P(V)}= M^{2S}_{P(V)} - M^{1S}_{P(V)}$ between the $1S$ and $2S$ state heavy pseudoscalar and vector mesons.
The mass gap $\Delta M_{P(V)}$ for pseudoscalar (vector) mesons in our LFQM is decomposed as
\begin{eqnarray} \label{eq:13}
\Delta M_{P(V)}^{} = \Delta M^{\rm Kin}_{P(V)} + \Delta M^{\rm Conf}_{P(V)} + \Delta M^{\rm Coul}_{P(V)} + \Delta M^{\rm Hyp}_{P(V)},
\end{eqnarray}
where we separate the four different contributions, i.e., $H_0$, $V_{\rm Conf}$, $V_{\rm Coul}$, and $V_{\rm Hyp}$, 
to the total mass gap for the taxonomical analysis in our numerical calculations. 
From the available experimental data for the $1S$ and $2S$ state heavy meson pairs, $(D,D^*)$, $(\eta_c, J/\Psi)$, and $(\eta_b, \Upsilon)$~\cite{PDG20},
we observe that the mass gaps between pseudoscalar mesons $(\Delta M_P)$ are greater than the corresponding mass gaps 
between vector mesons $(\Delta M_V)$, i.e., $\Delta M_P > \Delta M_V$.
In our LFQM calculation,  
$\Delta M^{\rm Kin + Conf + Coul}_{P}=\Delta M^{\rm  Kin + Conf + Coul}_{V}$
due to the usage of common $\beta$ parameters for both pseudoscalar and vector mesons of the same quark flavor contents
as shown in Eq.~(\ref{eq:12}), and thus
the mass gap is exclusively governed by the hyperfine interaction $V_{\rm Hyp}$ and can be readily obtained as
\begin{eqnarray} \label{eq:mass_gap}
\Delta M_P - \Delta M_V &=& \Delta M^{\rm Hyp}_P - \Delta M^{\rm Hyp}_V
\nonumber \\
&=& C \left(2 \sqrt{6}\sin 2\theta - \cos 2\theta \right),
\end{eqnarray}
where $C=16  \alpha_s \beta^3 / (9m_q m_{\bar{q}} \sqrt{\pi})$.

Equation~(\ref{eq:mass_gap}) combined with the relation $\Delta M_P > \Delta M_V$ provides a very important constraint on 
the mixing angle $\theta$.
It is evident that the pure ($\phi_{1S}, \phi_{2S}$) states with $\theta =0^\circ$ always leads to 
$\Delta M_P < \Delta M_V$, which shows that the introduction of the mixing is inevitable. 
Furthermore, one can find that the condition of $\Delta M_P > \Delta M_V$ gives the constraint,
\begin{equation} \label{eq:16}
\frac{1}{2}\cot^{-1} (2\sqrt{6}) < \theta < \frac{\pi}{4}.
\end{equation}
This concludes that the lower bound of the physical mixing angle, $\theta_c$, is determined as 
$\theta > \theta_c = \cot^{-1} (2\sqrt{6})/2\simeq 6^\circ$.

\begin{table*}[t]
	\begin{ruledtabular}
		\renewcommand{\arraystretch}{1.3}
		\caption{The constituent quark masses, potential parameters $(a,b,\alpha_s)$, and variational parameters 
		$\beta_{q\bar{q}}$ for the pure and mixed scenarios. 
		The quark masses, potential parameter $a$, and variational parameters $\beta$ are in the units of GeV, 
		while the string constant $b$ is in the unit of GeV$^2$.
		The strong coupling $\alpha_s$ is dimensionless and $q=u,d$.}
		\label{tab:parameter}
		\begin{tabular}{c|ccccc|cc|ccccccc}
			Mixing angle & $m_q$ & $m_s$  & $m_c$ & $m_b$ & $b$ & 
			$a$ & $\alpha_s$ &
			$\beta_{qc}$ & $\beta_{sc}$  & $\beta_{qb}$ & $\beta_{sb}$ & $\beta_{cc}$ & $\beta_{cb}$ & 
			$\beta_{bb}$\\ \hline
			Pure 	 $(\theta=0^\circ)$ & \multirow{2}{*}{0.22}     & \multirow{2}{*}{0.45}      & \multirow{2}{*}{1.68}      &  
			   \multirow{2}{*}{5.10} & \multirow{2}{*}{0.18} & $-0.538$ & 0.425 & 0.500    & 0.537     & 0.585      &  0.636 & 0.699 & 0.906 & 1.376\\ 
			Mixed  $(\theta=12^\circ)$&      		&  	 & &   &  & $-0.543$ & 0.433 &  0.424 & 0.455 & 0.495 & 0.538 & 0.592 & 0.767 & 1.167\\
		\end{tabular}
		\renewcommand{\arraystretch}{1}
	\end{ruledtabular}
\end{table*}

\subsection{\boldmath Model parameters}
\label{model-parameters}

As we have discussed in the previous subsection, the parameters of heavy mesons in the present model for $(1S, 2S)$ state mesons
include four quark masses ($m_q$, $m_s$, $m_c$, $m_b$) with $(q=u,d)$, seven variational HO parameters ($\beta_{qc}$, $\beta_{sc}$, 
$\beta_{qb}$, $\beta_{sb}$, $\beta_{cc}$, $\beta_{cb}$, $\beta_{bb}$), three potential parameters ($a$, $b$, $\alpha_s$), 
and the mixing angle $\theta$. 
The variational principle in Eq.~(\ref{eq:11}) leads to a constraint in the parameter space, which relates the strong coupling 
constant $\alpha_s$ and the other parameters i.e., $\alpha_s =\alpha_s(\theta, a, b, m_q, m_{\bar q},  \beta_{q{\bar q}})$.
This indicates that the variational parameters $\beta_{q{\bar q}}$ are automatically determined once other model parameters 
such as the quark masses, the strong coupling constant, the string tension, and the mixing angle are fixed.

In this study of heavy mesons, we take $m_q = 0.22$~GeV, $m_s = 0.45$~GeV, and the widely-used string tension 
$b=0.18$~GeV$^2$~\cite{GI85,ISGW89,SI95} as inputs, which were adopted in our previous LFQM analysis~\cite{CJ97,CJ99a,CJ09,Choi07} 
for $1S$ state mesons.
This leaves five parameters, i.e., ($m_c$, $m_b$, $a$, $\alpha_s$, $\theta$), to be determined.
In order to determine those five unknowns, we use two masses of the $1S$ state heavy mesons as inputs.
Among many possible choices of two input masses, we find that the use of the $(\eta_b, B^*)$ pair masses as inputs 
produces other meson masses well enough compared to the data.
Since we have only two equations $(M_{\eta_b}, M_{B^*})$ with five unknowns to be determined, we first try to find the best 
fit parameters for the pure $(1S, 2S)$ state case without mixing ($\theta=0^\circ$). 
In this case, we need to choose two input parameters from ($m_c$, $m_b$, $a$, $\alpha_s$).
Through our analyses with various combinations, we found that $m_c = 1.68$~GeV and $m_b = 5.10$~GeV give satisfactory results. 
We then obtain the remaining potential parameters, $a=-0.538$~GeV and $\alpha_s=0.425$, 
by solving Eq.~(\ref{eq:12}) for $(M^{1S}_{\eta_b}, M^{1S}_{B^*})$ using their measured values.
We also note that $V_{\rm Conf}$ and $V_{\rm Coul}$ are flavor- and scale-independent so that the confining potential constant $a$ 
and the strong coupling $\alpha_s$ are the same for all heavy mesons considered in this work.
Therefore, once $a$ and $\alpha_s$ are determined, the values of seven $\beta$ parameters are automatically computed and 
all the other meson masses are our predictions.

Using the same quark masses  $(m_q, m_s, m_c, m_b)$ and the string tension $b$ as in the $\theta=0^\circ$ case but taking 
into account of the two experimental constraints, $\Delta M^{\rm Hyp}_P > \Delta M^{\rm Hyp}_V$ and $f_{1S} > f_{2S}$,
we obtain the optimum value $\theta = 12^\circ$ of the mixing angle as well as other model parameters to cover both charm and 
bottom flavors of the heavy quark.%
\footnote{The mixing angle in general depends on the quark flavor contents of mesons. We found $\theta = 9.8^\circ$, $17.6^\circ$, 
and $13.9^\circ$ for $(D,D^*)$, $(\eta_c, J/\Psi)$, and $(\eta_b, \Upsilon)$, using the measured masses in PDG~\cite{PDG20}, i.e., 
$\Delta M_P - \Delta M_V = 62$~MeV, 64~MeV, and $37$~MeV, respectively.
However, the paucity of data does not allow us to estimate the mixing angles for the other mesons and, therefore, we analyze the quantities 
assuming $\theta= 12^\circ$, which was found to be fair in explaining the available data.}
This would be enough for verifying the mixing effects on the physical quantities of heavy mesons.

We summarize our best fits for the model parameters obtained for the mixed state ($\theta=12^\circ$) case in Table~\ref{tab:parameter}. 
For the comparison purpose of mixing effects, we also include the best fits for the model parameters obtained for the pure state 
($\theta=0^\circ$) case.
This shows that the values of $a$ and $\alpha_s$ are not significantly different in both cases, but the values of the $\beta$ parameters 
become smaller with mixing, which results in different meson properties.

\begin{figure}[t]
	\centering
	\includegraphics[width=0.9\columnwidth]{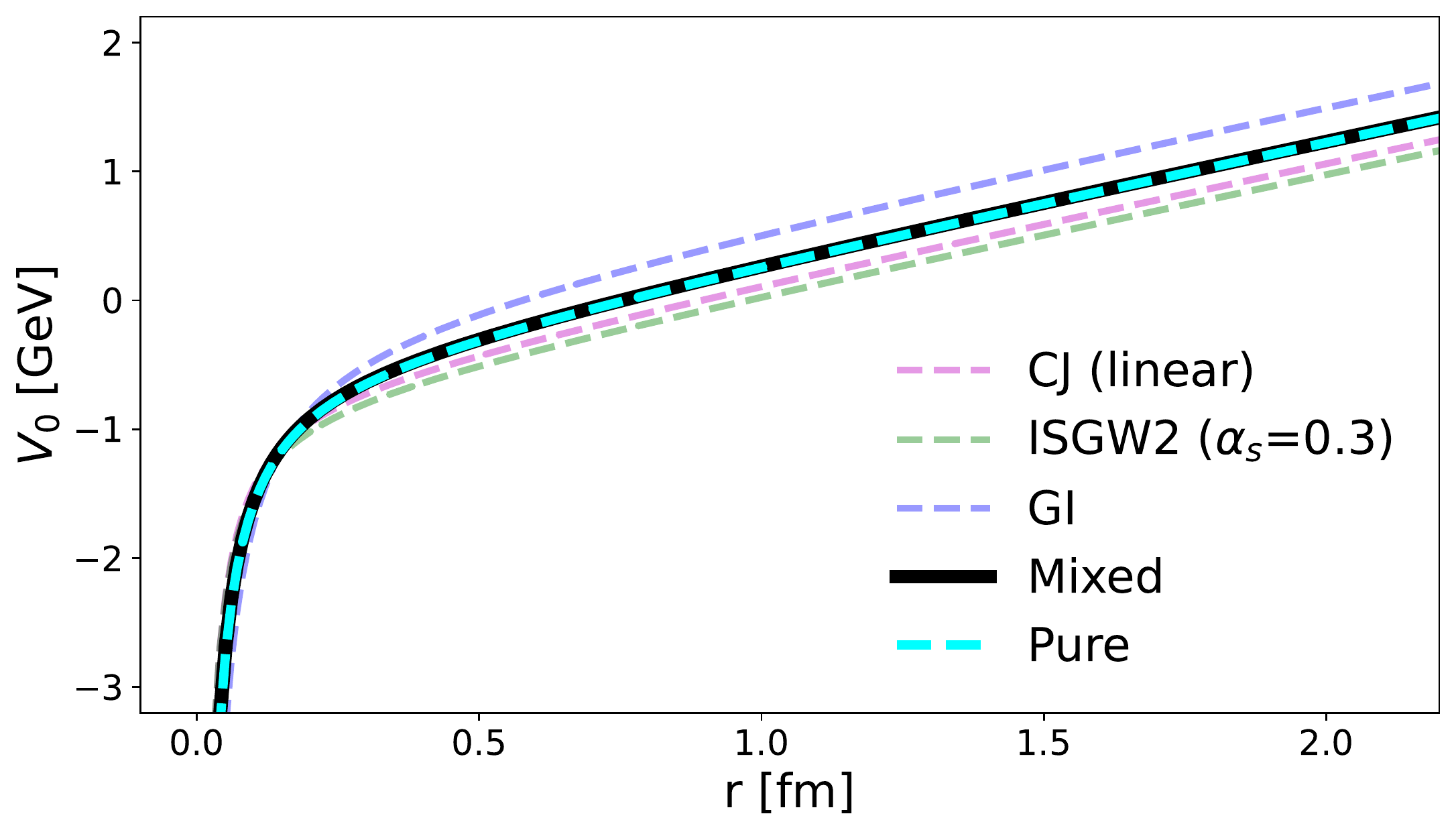}
	\caption{\label{fig:potential} 
	Central potentials in two different scenarios, the pure and mixed configurations. 
	No noticeable difference is observed between these two cases.
	The potentials of the CJ~\cite{Choi07}, ISGW2~\cite{SI95}, and GI~\cite{GI85} models are presented for comparison. 
	}
\end{figure}

With the model parameters determined as above, we can compare the central potential $V_0$ with other model calculations.
In Fig.~\ref{fig:potential}, we present the central potentials $V_0(r)$ up to $r\simeq 2$~fm for the pure and mixed configurations 
($\theta=0^\circ$ and $12^\circ$, respectively) and compare them with the well-known GI model~\cite{GI85} and ISGW2 model~\cite{SI95}.
We also plot the potential of the previous works of two of us in Refs.~\cite{CJ97,CJ99a,CJ09,Choi07} as the CJ model.
As one may expect from the similarities of the model parameters ($a$, $b$, $\alpha_s$), 
the central potentials obtained from the two different mixing scenarios are almost the same and
they are also quite comparable with the results from the GI and ISGW2 models as well as the CJ model.

\begin{figure*}[t]
	\centering
	\includegraphics[width=0.9\textwidth]{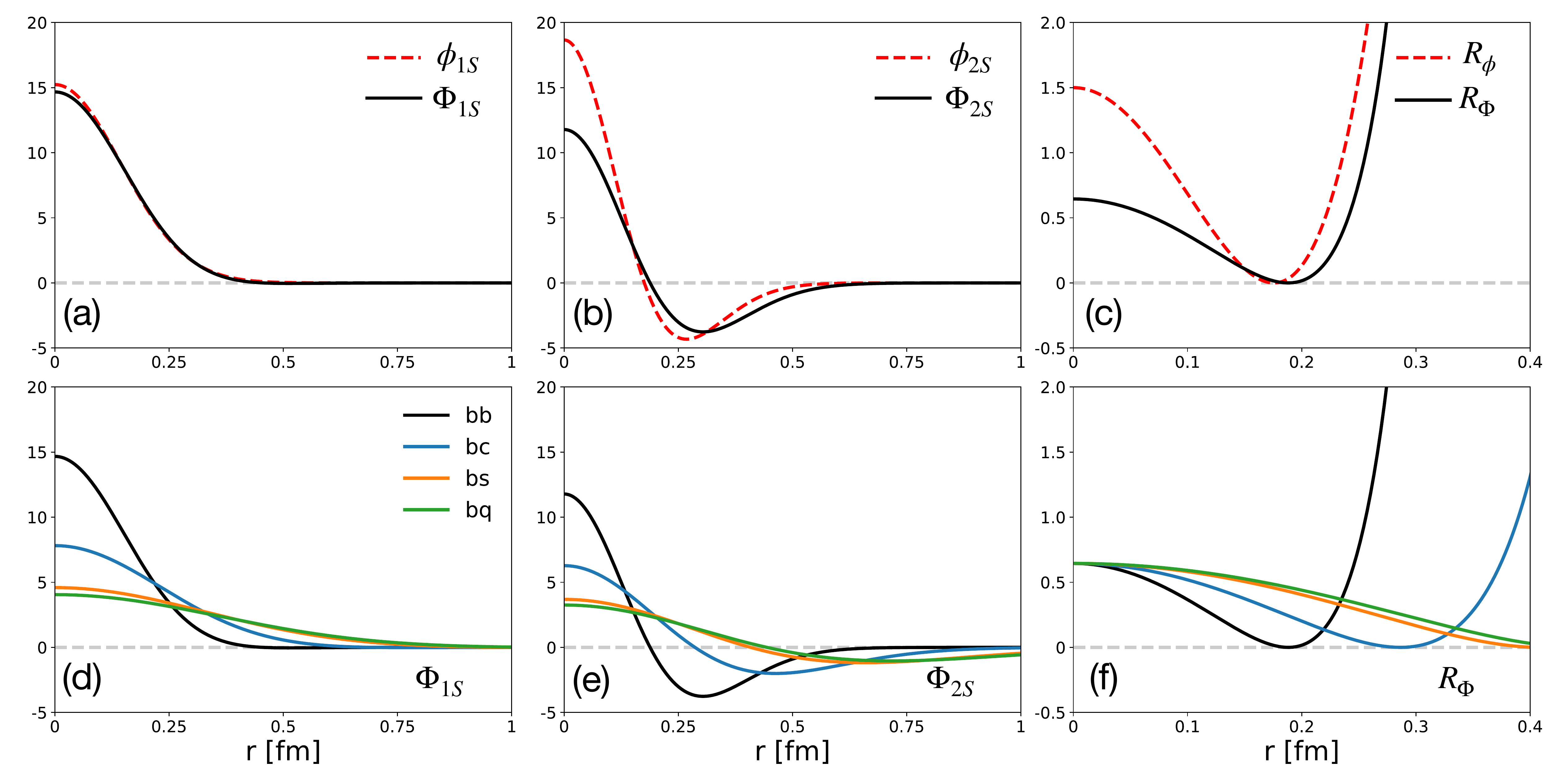}
	\caption{\label{fig:wavefunction} 
	(Upper panel) Radial wave functions in the pure and mixed configurations for the bottomonium $b\bar{b}$ states. 
	The mixing modifies the $2S$ wave function significantly, while the $1S$ wave function is barely unchanged. 
	The ratio is defined by $R_\phi=\phi_{2S}^2/\phi_{1S}^2$ and $R_\Phi=\Phi_{2S}^2/\Phi_{1S}^2$.
	(Lower panel) Radial wave functions in mixed scenario for various quark flavor contents.}
\end{figure*}

The mixing effects on the $1S$ and $2S$ state radial wave functions are shown in Fig.~\ref{fig:wavefunction} for the bottomonium 
($b\bar b$) case.
In Figs.~\ref{fig:wavefunction}(a) and \ref{fig:wavefunction}(b), we compare the two radial wave functions of the pure $\phi_{nS}$ 
states (dashed lines) and the mixed $\Phi_{nS}$ states (solid lines) for $n=1,2$.
Shown in Fig.~\ref{fig:wavefunction}(c) are the ratios $R_\phi=\phi^2_{2S}/\phi^2_{1S}$ (dashed line) and 
$R_\Phi=\Phi^2_{2S}/\Phi^2_{1S}$ (solid line).
This shows that the small mixing ($\theta=12^\circ$) significantly modifies the wave function of the $2S$ state, while the $1S$ radial 
wave function is barely modified.
The mixed state radial wave functions $\Phi_{nS}$ and their ratios for various heavy-heavy ($b{\bar b}$, $b{\bar c}$)
and heavy-light ($b{\bar s}$, $b{\bar q}$) with ($q=u$, $d$) quark states are also given in Figs.~\ref{fig:wavefunction}(d)-\ref{fig:wavefunction}(f).
Since the range of the radial wave function is inversely proportional to the value of the $\beta$ parameter, 
it is quite natural to observe that the wave function of bottomonium state is narrower than the other states.
We also note that the value of the radial wave function at the origin ($r=0$) is proportional to the $\beta$ parameter. 
As shown in Figs.~\ref{fig:wavefunction}(d) and~\ref{fig:wavefunction}(e), the bottomonium wave function at the origin is the highest among those 
of meson wave functions for both the $1S$ or $2S$ state.
However, the ratio $R_\Phi$ at the origin takes the same value independent of the quark flavor contents once the mixing angle is fixed as
can be seen in Fig.~\ref{fig:wavefunction}(f).

\subsection{\boldmath Mass spectra}

\begin{table*}[t]
	\begin{ruledtabular}
		\renewcommand{\arraystretch}{1.2}
		\caption{
		Mass spectra of the $1S$ and $2S$ state heavy mesons in the units of MeV.}
		\label{tab:masses}
		\begin{tabular}{lccccc}
        State     & Pure ($\theta=0^\circ$)  & Mixed ($\theta=12^\circ$)  & Expt.~\cite{PDG20} & GI~\cite{GI85} & RQM~\cite{EFG02b,EFG09b} \\ \hline 
        $D(1S)$   & 1731 & 1745 & 1869.66(05) & 1880 & 1871\\
        $D(2S)$   & 2282 & 2432 & 2549(19) & 2580 & 2581\\ 
        $D^*(1S)$ & 2020 & 2017 & 2010.26(05) & 2040 & 2010\\
        $D^*(2S)$ & 2714 & 2608 & 2627(10) & 2640 &2632\\ \hline
        $D_s(1S)$   & 1938 & 1946 & 1968.35(7) & 1980 & 1969\\
        $D_s(2S)$   & 2546 &2600 & 2591(6)\footnote{From the recent observation by the LHCb Collaboration~\cite{LHCb-20d}.}  & 2670 & 2688\\ 
        $D^*_s(1S)$ & 2113 & 2111 & 2112.2(4)  & 2130 & 2111\\
        $D^*_s(2S)$ & 2798 & 2706 & 2714(5) & 2730 & 2731\\ \hline 
        $\eta_c(1S)$& 2987 & 2990 & 2983.9(4) & 2970 & 2979\\
        $\eta_c(2S)$& 3627 & 3608 & 3637.5(1.1) & 3620 & 3588\\
        $J/\Psi(1S)$& 3090 & 3087 & 3096.900(6) & 3100 & 3096\\
        $\Psi(2S)$& 3781  & 3670 & 3686.10(6) & 3680 & 3686\\ \hline 
        $B(1S)$   & 5174 & 5182 & 5279.34(12)    & 5310 & 5280\\
        $B(2S)$   & 5740 & 5794 & ---   & 5900 & 5890\\ 
        $B^*(1S)$ & 5325 & 5325 & 5324.70(21)  & 5370 & 5326\\
        $B^*(2S)$ & 5968 & 5886  & ---  & 5930 & 5906\\ \hline 
        $B_s(1S)$   & 5325 & 5330 & 5366.88(14)  & 5390 & 5372\\
        $B_s(2S)$   & 5924 & 5928 & --- & 5980 & 5976\\ 
        $B^*_s(1S)$ & 5421 & 5418 & $5415.4^{+1.8}_{-1.5}$   & 5450 & 5414\\
        $B^*_s(2S)$ & 6067 & 5987 & --- & 6010 & 5992\\ \hline 
        $B_c(1S)$   & 6269 & 6270 & 6274.47(32)   & 6270 & 6270\\
        $B_c(2S)$   & 6948 & 6885 & --- & 6850 & 6835\\ 
        $B^*_c(1S)$ & 6270 & 6340 & --- & 6340 & 6332\\
        $B^*_c(2S)$ & 7059 & 6930 & --- & 6890 & 6881\\ \hline 
        $\eta_b(1S)$& 9399 & 9399 & 9398.7(2.0) & 9400 & 9400\\
        $\eta_b(2S)$& 10249 & 10123 & 9999(4) & 9980 & 9993\\
        $\Upsilon(1S)$& 9485 & 9480 & 9460.30(26) & 9460 & 9460\\
        $\Upsilon(2S)$& 10377 & 10175 & 10023.26(31) & 10000 & 10023 \\
		\end{tabular}
		\renewcommand{\arraystretch}{1}
	\end{ruledtabular}
\end{table*}

In this subsection, we compute the mass spectra of the $1S$ and $2S$ state heavy mesons. 
With the parameters given in Table~\ref{tab:parameter}, the mass formulas in Eq.~ (\ref{eq:12}) are used to obtain the mass spectra.
The meson masses obtained for $\theta=0^\circ$ and $\theta = 12^\circ$ cases are summarized in Table~\ref{tab:masses}. 
As discussed before, all the meson masses apart from the two inputs ($M^{1S}_{\eta_b}, M^{1S}_{B^*}$) are our predictions.
For comparison, we also list the experimental data of Ref.~\cite{PDG20} and the predictions of the GI model~\cite{GI85}
and those of the relativistic quark model (RQM) in Refs.~\cite{EFG02b,EFG09b}.
Our predictions obtained from both mixing scenarios are found to be overall in a good agreement with the experimental data~\cite{PDG20}.
However, there are some delicate but important mixing effects in mass spectra.
While there are no big differences in the predicted masses for the $1S$ state mesons between the pure and mixed scenarios, 
the predictions for the $2S$ state mesons obtained from the mixed case agree better with the experimental data.
This also can be seen by performing $\chi^2$ analysis, which gives $\chi^2 =0.009$ for the mixed scenario and $\chi^2 =0.024$ 
for the pure one.%
\footnote{The $\chi^2$ is computed as $\chi^2= \sum_i \left[(O_i-E_i)^2/E_i^2\right]$, 
where $O_i$ and $E_i$ are the experimental data and theoretical prediction, respectively.}

In particular, our result, $M_{D_s(2S)}=2600$ MeV for the $2S$ state of the $D_s$ meson with $\theta=12^\circ$ is very close 
to the observed mass of the recently discovered $D_{s0}(2590)^+$ with $J^P = 0^-$~\cite{LHCb-20d}.
This supports the interpretation of the observed $D_{s0}(2590)^+$ state as a standard quark-antiquark radial excitation of the 
$D_s^+$ meson as claimed by the LHCb Collaboration~\cite{LHCb-20d}.
For making a definite conclusion on the structure of the $D_{s0}(2590)^+$, however, we still need more detailed and precise
experimental studies on various properties of the $D_{s0}(2590)^+$.

The LHCb Collaboration~\cite{LHCb-15a} also reported traces of $B_J(5840)$ and $B_J(5960)$, and confirmed
the observation of the $B_J(5960)$ by the CDF Collaboration~\cite{CDF-13}.
Although their existence as resonances awaits confirmation and their quantum numbers are yet to be identified,
these states are suggested as the $2S$ states of $B$ and $B^*$ mesons in Ref.~\cite{LHCb-15a}.
However, if these are the $B(2S)$ and $B^*(2S)$ states, then it violates the mass hierarchy by giving $\Delta M_P < \Delta M_V$
as $\Delta M_P \approx 561$~MeV and $\Delta M_V \approx 635$~MeV. 
This is in contradiction with our predictions on the mass gaps, $\Delta M_P \approx 612$~MeV and $\Delta M_V \approx 561$~MeV, 
which observe the relation $\Delta M_P > \Delta M_V$.
Therefore, verifying the $B(2S)$ and $B^*(2S)$ states is crucial to understand the structure of the radially excited heavy meson states.
Experimental searches for these states in $B$, $B^*$, $B_s$ and $B_s^*$ are thus highly anticipated.

\begin{figure*}[t]
	\centering
	\includegraphics[width=0.95\textwidth]{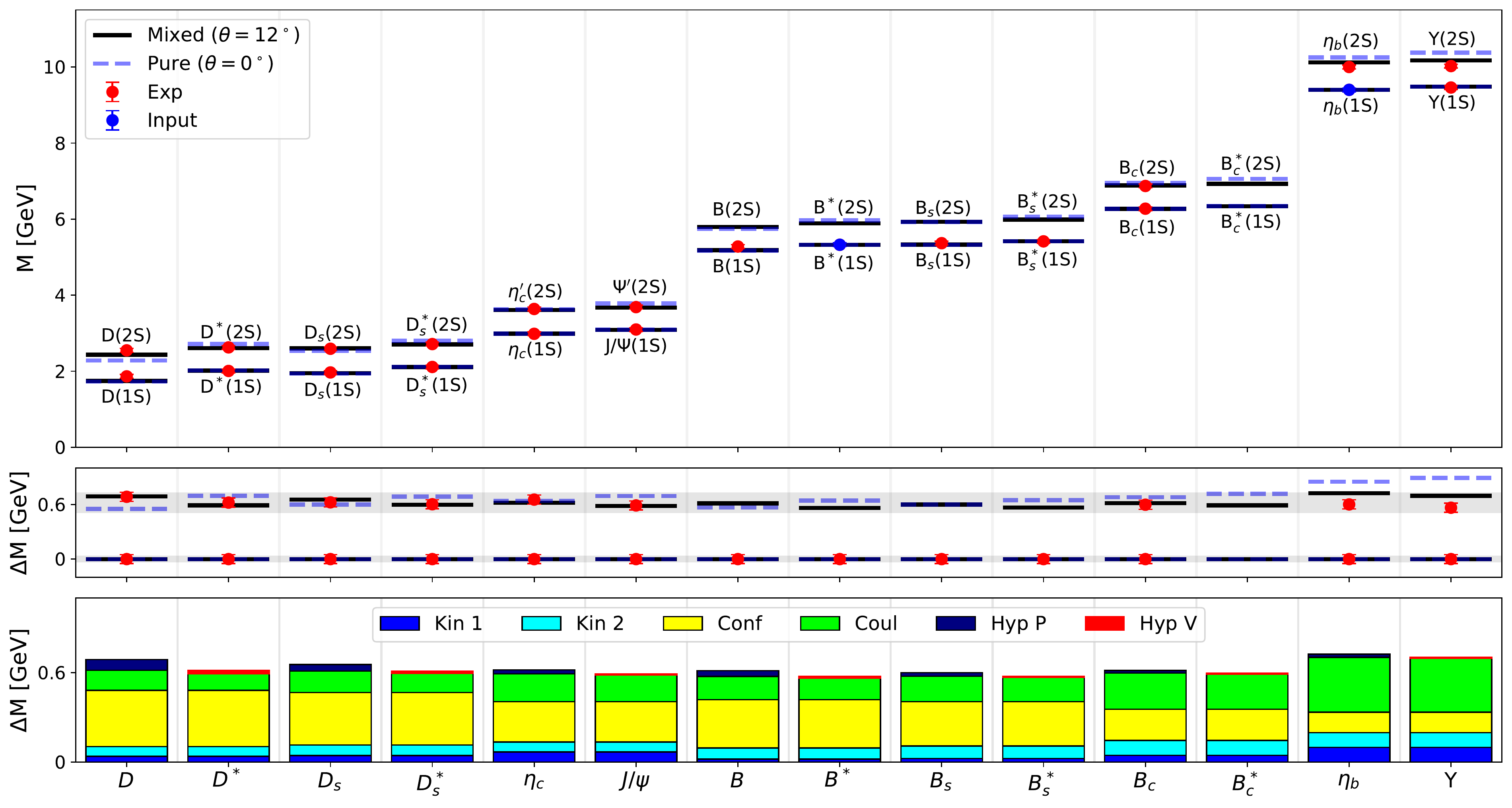}
	\caption{\label{fig:mass} (Upper panel) 
	         Mass spectra of $1S$ and $2S$ state heavy mesons in the pure and mixed configurations. 
		The experimental data are taken from Ref.~\cite{PDG20} and the recent observation of the LHCb Collaboration~\cite{LHCb-20d}.  
		(Middle panel) The mass gap between the $1S$ and $2S$ heavy mesons. 
		The masses of all the $2S$ states are given relative to the $1S$ state masses.
		The mass gap is observed to be around 600~MeV regardless of the quark flavor contents. 
		(Lower panel) The computed component of the mass gap.
		When the quark and antiquark have different masses, the contribution of the heavier (lighter) quark in $H_0$ is denoted as Kin 1 (Kin 2).}
\end{figure*}

The top panel of Fig.~\ref{fig:mass} shows the mass spectra of the $1S$ and $2S$ state heavy mesons.
The middle panel of Fig.~\ref{fig:mass} represents the mass gaps between the $1S$ and $2S$ state mesons
and the four different contributions to this mass gap ($\Delta M^{\rm Kin}$, $\Delta M^{\rm Conf}$, $\Delta M^{\rm Coul}$,  
$\Delta M^{\rm Hyp}$) are depicted in the bottom panel of Fig.~\ref{fig:mass}.
The dashed and solid lines in the upper and middle panels represent our results obtained with the pure ($\theta=0^\circ$) and 
mixed ($\theta=12^\circ$) cases, respectively. 
The decomposition of $\Delta M$ shown in the bottom panel of Fig.~\ref{fig:mass} is for the mixed case.
We also note in this taxonomical analysis that the contributions of the heavier and lighter quarks in the kinetic energy part are 
further separated and denoted as `Kin 1' and `Kin 2,' respectively, when the quark contents are different.
As one can see from the mass gap $\Delta M$, the observed mass gap relation, $\Delta M_P > \Delta M_V$, cannot be realized 
without introducing the mixing angle.
It is also interesting to see from the available data that the mass gaps between the $1S$ and $2S$ states are around
600~MeV, and the values are almost flavor-independent.
Similar mass gap is also observed for the radially excited states of baryons with various flavors~\cite{ANHT20b}.

While the mass gap $\Delta M$ seems almost flavor-independent as shown in the middle panel of Fig.~\ref{fig:mass}, 
the four different contributions, ($\Delta M_{\rm Kin}$, $\Delta M_{\rm Conf}$, $\Delta M_{\rm Coul}$, $\Delta M_{\rm Hyp}$),
which make up $\Delta M$, are flavor-dependent as one can see from the bottom panel of Fig.~\ref{fig:mass}.
For instance, comparing the Coulomb and confinement interactions, one can easily find from Eq.~(\ref{eq:12}) that
$\Delta M_{\rm Conf} \propto \beta^{-1}$ while $\Delta M_{\rm Coul} \propto \beta$.
This relation provides an intuitive explanation for the observation that the confinement is dominant at large distances, while the Coulomb 
interaction arising from the one-gluon exchange dominates at short distances.
This tendency can be clearly seen in the bottom panel of Fig.~\ref{fig:mass}.
By comparing the mass gap components of $(\eta_c, \eta_b)$ or $(J/\psi, \Upsilon)$, one can see that the green area for 
$\Delta M_{\rm Coul}$ becomes larger for bottom quark systems.
We also find that the kinetic energy is one of the important components in the mass gap.
In particular, for mesons with different quark and antiquark masses such as $B$ and $D$ mesons, we find that the light quark (Kin2) 
gains more kinetic energy than the heavy quark (Kin1) as one may expect intuitively.

\section{Applications} 
\label{sec:applications}

With the model parameters fixed by mass spectrum, we predict various properties of the $1S$ and $2S$ state heavy mesons in this section.
The standard LFQM adopting the spin-orbit wave functions and the polarization vectors of a vector meson defined in 
Eqs.~(\ref{eq:6}) and (\ref{eq:7}) is based on the requirement of all constituents being on their respective mass shell (i.e., $M \to M_0$). 
This on-mass-shell condition of quark and antiquark is completely different from the manifestly covariant models, which 
allow the quark and antiquark to be off-mass-shell allowing $M \neq M_0$. 
For instance, the invariant mass $M_0$ included in the polarization vector $\epsilon(0)$ of Eq.~(\ref{eq:7}) in the standard LFQM needs to be 
replaced by the physical mass $M$ in the manifestly covariant model.

The complications coming from the binding energy issue do not appear in the analysis of meson mass spectra since 
only the radial wave functions are needed. 
However, it does matter for the calculations of other physical observables such as the decay constants and form factors, which we will discuss below.
In the previous works of Refs.~\cite{CJ13,CJ14,CJ17,Choi21,Choi21b} for the decay constants, DAs
for pseudoscalar and vector mesons, and weak transition form factors between two pseudoscalar mesons, 
it was shown that the self-consistent LFQM description of those physical observables can be achieved if and only 
if every physical mass $M$ appeared in the matrix elements is replaced by the invariant mass $M_0$. 
In other words, the replacement of the physical mass $M$ in the integrand of the amplitude by the invariant mass $M_0$ 
(denoted by CJ-scheme for convenience) results in the physical observables that are independent of the current 
components and polarizations used in the analysis.
This $M \to M_0$ mapping is indeed proven to be an effective way of including the treacherous points such as the
light-front zero modes and the instantaneous contributions.
As the comprehensive and rigorous analysis of decay constants and DAs for pseudoscalar and vector mesons can be found in 
Refs.~\cite{CJ13,CJ14,CJ17,Choi21}, here we just summarize the final theoretical results for those physical quantities for completeness.

\subsection{Decay constants}

The decay constants of a pseudoscalar meson $P$ and a vector meson $V$ with a four-momentum $P^\mu$ and 
a mass $M$ are defined by
\begin{eqnarray}\label{eq:17}
	\bra{0} \bar{q} \gamma^\mu \gamma_5 q \ket{P} &=& i f_P P^\mu, \nonumber\\
	\bra{0} \bar{q} \gamma^\mu q \ket{V(P,\lambda)} &=& f_V M \epsilon^\mu(\lambda),
\end{eqnarray}
as $f_P$ and $f_V$, respectively, where $\epsilon^\mu (\lambda)$ is the polarization vector of a vector meson given by Eq.~(\ref{eq:7}).
For the case of pseudoscalar mesons, it has been explicitly shown that the decay constants of 
the pure $1S$ state mesons obtained from the plus ($\mu=+$) and minus ($\mu=-$) components of the currents 
are exactly the same~\cite{Choi21}.
In the present work, we extend it to the cases of mixed $1S$ and $2S$ states.
Denoting $f^{(\pm)}_P$ obtained from the plus and minus components of the currents, 
the explicit forms of $f^{(\pm)}_{P}$ are given by~\cite{Choi21} 
\begin{eqnarray}\label{eq:18}
	f_P^{(\pm)} = \sqrt{6} \int^1_0 dx \int \frac{d^2 \mathbf{k}_\bot}{(2\pi)^3}  
	\frac{ {\Phi}(x, \mathbf{k}_\bot) }{\sqrt{\mathcal{A}^2 + \mathbf{k}_\bot^2}} ~\mathcal{O}^{(\pm)}_{P},
\end{eqnarray}
where $\mathcal{A} =  (1-x)  m_q+  x m_{\bar{q}}$ and 
\begin{eqnarray} \label{eq:19}
\mathcal{O}^+_{P} &=& \mathcal{A},\nonumber\\
\mathcal{O}^-_{P} &=& \frac{{\bf k}^2_\perp \mathcal{A'} + m_q m_{\bar q}\mathcal{A}}{ x(1-x) M^2_0},
\end{eqnarray}
with $\mathcal{A'}=\mathcal{A}(m_q \leftrightarrow m_{\bar q})$. 
Here, ${\Phi}(x, \mathbf{k}_\bot)$ denotes the wave functions (${\Phi}_{1S}$, ${\Phi}_{2S}$) defined in Eq.~(\ref{eq:8}) 
for $(1S, 2S)$ state decay constants.

For the vector meson case, it was also explicitly shown that the decay constants for the pure $1S$ state
mesons obtained from the plus ($\mu=+$) component of the currents with the longitudinal polarization $\epsilon(0)$ 
and the perpendicular ($\mu=\perp$) components of the currents with the transverse polarizations $\epsilon(\pm)$  
are exactly the same to each other~\cite{CJ13}.
Denoting $f_V$ obtained from the plus and perpendicular components of the currents as $f^{(+)}_V$ and $f^{(\perp)}_V$, 
respectively, their explicit forms read~\cite{CJ13} 
\begin{eqnarray}\label{eq:20}
	f_V^{(+,\perp)} &=& \sqrt{6} \int^1_0 dx \int\frac{d^2 \mathbf{k}_\bot}{(2\pi)^3}   
\frac{ {\Phi}(x, \mathbf{k}_\bot) }{\sqrt{\mathcal{A}^2 + \mathbf{k}_\bot^2}} 
\mathcal{O}^{(+,\perp)}_V, \quad 
\end{eqnarray}
where 
\begin{eqnarray} \label{eq:21}
\mathcal{O}^+_{V} &=& \mathcal{A} + \frac{2 \mathbf{k}_\bot^2}{D_{LF}},\nonumber\\
\mathcal{O}^\perp_{V} &=& \frac{1}{M_0}
\left[ \frac{{\bf k}^2_\perp + {\cal A}^2}{2x(1-x)} -{\bf k}^2_\perp
  + \frac{ (m_q + m_{\bar{q}})}{D_{\rm LF}}{\bf k}^2_\perp
 \right],
\end{eqnarray}
and $D_{\rm LF}= M_0 + m_q + m_{\bar{q}}$.
In Ref.~\cite{LLMV18}, both $f^{(+)}_V$ and $f^{(\perp)}_V$ were computed in the BLFQ approach, but the two results 
are found to depend on the adopted component of the current, which was ascribed to the measure of rotational symmetry 
violation of the model. 
In our LFQM calculation, by using the CJ-scheme, however, we could confirm numerically that $f_P \equiv f_P^{(+)} =f_P^{(-)}$ 
and $f_V\equiv f_V^{(+)} =f_V^{(\perp)}$ for both $(1S,2S)$ state mesons.

\subsection{Twist-2 distribution amplitudes}

The twist-2 quark DAs, $\phi^{\rm tw-2}_{P(V)}(x)$, for pseudoscalar and vector mesons are related with
the decay constants obtained from the plus component of the currents through~\cite{CJ07}
\begin{equation} \label{eq:22}
\int^1_0 \phi^{\rm tw-2}_{P(V)}(x,\mu) dx = \frac{f^{(+)}_{P(V)} }{2\sqrt{6}},
\end{equation}
where $\phi^{\rm tw-2}_{P(V)}(x,\mu)$ is obtained by the ${\bf k}_\perp$ integration of the LF wave function up to the transverse momentum scale $\mu$. Here, $\mu (\geq |{\bf k}_\perp|)$ can be regarded as the energy scale that separates the perturbative and non-perturbative regimes.
The twist-2 DA then describes the probability amplitudes to find the hadron in a state with a minimum number of 
Fock constituents and small transverse momentum separation.
While the typical transverse momentum cutoff for the light meson sectors~\cite{CJ13,CJ14,CJ17} was estimated 
as $\mu\approx 1$ GeV, the values of the scale $\mu$ for the heavy meson sectors appear shifted to the larger values as we discuss in the numerical results of Sec.\ref{numerical-DA}.

The normalized quark DA is defined as ${\tilde\phi}^{\rm tw-2}_{P(V)}(x,\mu)=(2\sqrt{6}/f^{(+)}_{P(V)})\phi^{\rm tw-2}_{P(V)}(x,\mu)$
so that
\begin{equation} \label{eq:23}
	\int_{0}^{1} {\tilde\phi}^{\rm tw-2}_{P(V)}(x,\mu) dx = 1.
\end{equation}
The quark DAs can be usually expanded in Gegenbaur polynomials $C_n^{3/2}$
as ${\tilde\phi}(x,\mu) = {\tilde\phi}_{\rm as}(x) \left[ 1 + \sum^{\infty}_{n=1} a_n(\mu) C_n^{3/2}(\xi) \right]$,
where the Gegenbaur moments $a_n(\mu)$ gauge the deviation of the DAs from the asymptotic one ${\tilde\phi}_{\rm as}(x)=6x(1-x)$.
Alternatively,  one can define the expectation value of the longitudinal momentum, i.e., the $\xi=x-(1-x)=2x-1$ moments
defined as~\cite{CJ07}
\begin{equation} \label{eq:xi}
\expval{\xi^n} = \int_{0}^{1} dx\ \xi^n\ {\tilde\phi}^{\rm tw-2}_{P(V)}(x,\mu),
\end{equation}
which are closely related with the Gegenbaur moments $a_n(\mu)$. The explicit relations
between $\expval{\xi^n}$ and $a_n(\mu)$ can be found, for example, in Ref.~\cite{CJ07}.

\begin{figure*}[t]
	\centering
	\includegraphics[width=0.95\textwidth]{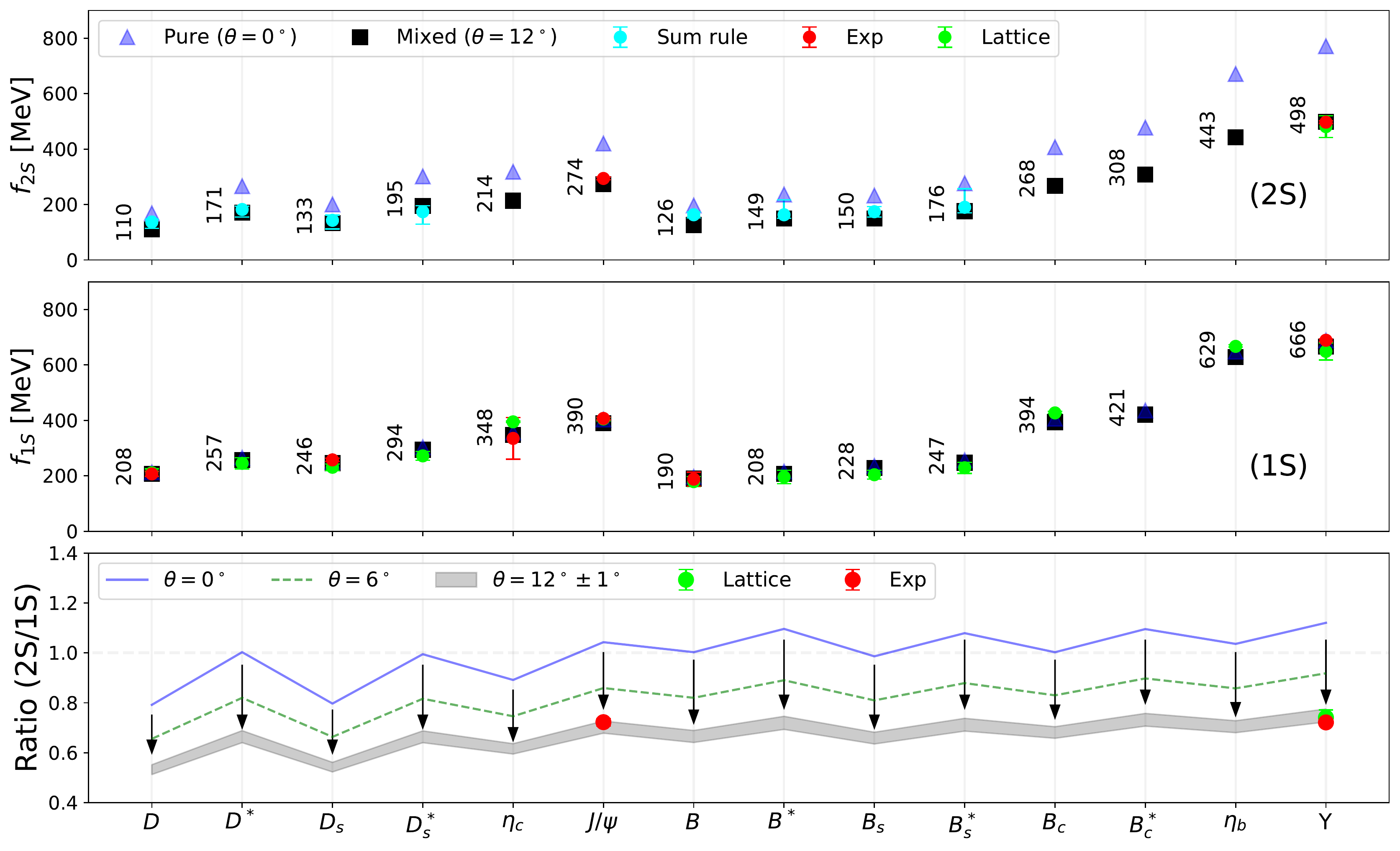}
	\caption{\label{fig:decay-constant} 
	(Upper and middle panels) Comparison of the predicted decay constants of the $1S$ and $2S$ state heavy mesons 
	with experimental data~\cite{PDG20} and lattice simulations~\cite{HPQCD-12a,HPQCD-10,HPQCD-12,HPQCD-14}. 
	For the $2S$ state, we include the QCD sum rule result of Ref.~\cite{GKPR14}. 
	(Lower panel) The ratio $R_f=f_{2S}/ f_{1S}$ compared to the experimental data and lattice simulations. 
	}
\end{figure*}

\subsection{Electromagnetic form factors and charge radii}

We also compute the electromagnetic form factors of heavy pseudoscalar mesons as well as their charge radii.
Our calculation is carried out by using the Drell-Yan-West frame $(q^+=0)$ with $\mathbf{q}_\perp^2 =Q^2 = -q^2$.
The electromagnetic form factor of the pseudoscalar meson can be expressed for the ``$+$'' component of the current $J^\mu$ as~\cite{CJ97}
\begin{equation}
	F(Q^2) = e_q I^+(Q^2, m_q, m_{\bar{q}}) + e_{\bar{q}} I^+(Q^2, m_{\bar{q}},m_q), \quad \quad 
\end{equation}
where $e_q (e_{\bar{q}})$ is  the electric charge of quark (antiquark), and
\begin{eqnarray}
	I^+(Q^2, m_q, m_{\bar{q}}) &=& \int^1_0 dx \int\frac{ d^2 \mathbf{k}_\perp}{2(2\pi)^3}
	\, \Phi(x,\mathbf{k}_\perp) \, \Phi^*(x,\mathbf{k}^\prime_\perp)  
\nonumber\\	& & \mbox{} \qquad \times 
        \frac{\mathcal{A}^2 + \mathbf{k}_\perp \cdot \mathbf{k}^\prime_\perp}{\sqrt{\mathcal{A}^2 + \mathbf{k}_\perp^2} 
	\sqrt{\mathcal{A}^2 + \mathbf{k}_\perp^{\prime 2}}},
\end{eqnarray}
where $\mathbf{k}^\prime_\perp = \mathbf{k}_\perp + (1-x) \mathbf{q}_\perp$.
The electromagnetic form factor is normalized as $F(0) = e_q + e_{\bar{q}}$, and the charge radius of the meson is calculated by
\begin{eqnarray}\label{eq:24}
	\Braket{ r^2 } = -6 \frac{dF(Q^2)}{dQ^2}\biggl|_{Q^2=0}.
\end{eqnarray}

\section{Results and discussion} \label{sec:results}

\subsection{Decay constants}

In Fig.~\ref{fig:decay-constant}, we present our numerical results for the decay constants of the $1S$ (middle panel) and 
$2S$ (upper panel) state heavy mesons.
The results obtained with the pure $(\theta=0^\circ)$ and mixed $(\theta=12^\circ)$ wave functions are represented 
by triangles and squares, respectively.
For comparison, we show the available experimental data~\cite{PDG20} and other theoretical predictions from
lattice simulations~\cite{BBLLMMR98,HPQCD-12a,HPQCD-10,HPQCD-12,HPQCD-14} and the QCD sum rules~\cite{GKPR14}.

\begin{table*}[t]
	\begin{ruledtabular}
		\renewcommand{\arraystretch}{1.0}
		\caption{Decay constants of the $1S$ and $2S$ heavy-light mesons in the units of MeV.}
		\label{tab:constant_a}
		\begin{tabular}{lcccccccc}
			& $f_{D(1S)}$ 		  &	$f_{D^*(1S)}$ 		& $f_{D_s(1S)}$		&  $f_{D^*_s(1S)}$	& $f_{B(1S)}$		&  $f_{B^*(1S)}$		& $f_{B_s(1S)}$		
			& $f_{B_s^*(1S)}$	 \vspace{3pt} \\ \hline  
			Pure	 ($\theta = 0^\circ$) &  212	& 265	& 251	& 303 	& 196 	& 215 	& 235	& 256 	 \\ 
			Mixed ($\theta=12^\circ$)	  &  208	& 257	& 246	& 294	& 190 	& 208 	& 228	& 247 	 \\   
			Expt.~\cite{PDG20}		 & $206.7\pm 8.9$ &  ---& $257.5\pm 6.1$& --- & $188\pm 25$ & --- & --- & ---  \\
			Lattice~\cite{BBLLMMR98} & $211\pm 14 $ & $245\pm20 $ & $231 \pm 12$ & $272\pm 16$ & $179\pm 18$ & $196\pm 24 $ & $204\pm 16$ & $229\pm20$ \\ 
			Sum rules~\cite{Wang15} & $208\pm10$ & $263\pm21$ & $240\pm 10$ & $308\pm 21$ & $194\pm15$ & $213\pm18$ & $231\pm 16$ & $255\pm 19$ \\ 
			BS~\cite{CKWN04} & $230\pm 25$ & $340\pm 23$ & $248\pm 27$ & $375\pm 24$ & $196\pm 29$ & $238\pm 18$ & $216\pm32$ & $272\pm 20$ \\
			BS~\cite{BCDGPR18} & 223(11) & --- & 242(8) & --- & 201(18) & --- & 253(17) & --- \\
			LFQM (CJ)~\cite{CJ09} &  197	& 239	& 233	& 274 	& 171 	& 186 	& 205	& 220 		\\
			LFQM (CJ2)~\cite{CJLR15} &  208	& 230	& 232	& 260 	& 181 	& 185 	& 205	& 216 		\\  
			LFQM~\cite{DDJC19} & 197  & 230 & 219 & 253 & 163 & 172 & 184 & 194 \\
			RQM~\cite{EFG06} & 234 & 310 & 268 & 315 & 189 & 219 & 218 & 251 \\ \hline 
			& $f_{D(2S)}$ 		  &	$f_{D^*(2S)}$ 		& $f_{D_s(2S)}$		&  $f_{D^*_s(2S)}$	& $f_{B(2S)}$		&  $f_{B^*(2S)}$		& $f_{B_s(2S)}$		
			& $f_{B_s^*(2S)}$	  \\ \hline  
			Pure	($\theta=0^\circ$)   &  168	& 266	& 200	& 301 	& 197 	& 236	& 232	& 276 	 \\ 
			Mixed($\theta=12^\circ$)	 &  110	& 171	& 133	& 195 	& 126 	& 149 	& 150	& 176 	 \\    
			Sum rules, set I~\cite{GKPR14} & $137^{+10}_{-23}$	&  $182^{+12}_{-27}$ & $143^{+19}_{-31}$		& $174^{+22}_{-45}$	 	& $163^{+10}_{-11}$	  	& $163^{+54}_{-13}$	  	& $174^{+19}_{-19}$	 	& $190^{+67}_{-20}$	   \\
			Sum rules, set II~\cite{GKPR14} & $138^{+10}_{-22}$	&  $183^{+13}_{-24}$ & $146^{+12}_{-36}$		& $178^{+30}_{-39}$	 	& $166^{+9}_{-10}$	  	& $165^{+46}_{-12}$	  	& $178^{+19}_{-17}$	 	& $194^{+57}_{-18}$\\
			RQM~\cite{SPV14} & 292.14 & 293.38 & ---  & ---  & --- & ---  & --- &  --- \\
		\end{tabular}
		\renewcommand{\arraystretch}{1}
	\end{ruledtabular}
\end{table*}

While the decay constants of the $1S$ state mesons are rather robust against the change of the mixing angle,
those of the $2S$ state mesons are shown to be quite sensitive to the mixing angle, and clearly a better description is achieved 
thanks to the mixing effects.
For instance, our predictions for the heavy quarkonia obtained with the mixed wave functions are
$(f_{J/\Psi(1S)},f_{J/\Psi(2S)}) = (390, 274)$~MeV and $(f_{\Upsilon(1S)},f_{\Upsilon(2S)}) = (666, 498)$~MeV.
These values not only satisfy the hierarchy of $f_{1S}>f_{2S}$ but also consistent with the experimental data~\cite{PDG20}:
$(f^{\rm Expt.}_{J/\Psi(1S)},f^{\rm Expt.}_{J/\Psi(2S)}) = \bm{(} 407(5), 294(5) \bm{)}$~MeV and 
$(f^{\rm Expt.}_{\Upsilon(1S)},f^{\rm Expt.}_{\Upsilon(2S)})= \bm{(} 689(5), 497(5) \bm{)}$~MeV.
The full numerical results of our LFQM compared with the available experimental data~\cite{PDG20} as well as other theoretical model
predictions~\cite{CJ09,CJLR15,DDJC19,LLCLV21,Wang15,CKWN04,BCDGPR18,EFG06,CG90b,BDKMS13,LMV17}
are summarized in Tables~\ref{tab:constant_a} and~\ref{tab:constant_b}.

\begin{table*}[t]
	\begin{ruledtabular}
		\renewcommand{\arraystretch}{1.0}
		\caption{Decay constants of the $1S$ and $2S$ state $B_c$ and heavy quarkonia in the units of MeV.}
		\label{tab:constant_b}
		\begin{tabular}{lcccccc}
			 & $f_{\eta_c(1S)}$	& $f_{J/\Psi(1S)}$  & $f_{B_c(1S)}$		& $f_{B_c^*(1S)}$	 & $f_{\eta_b(1S)}$  & $f_{\Upsilon(1S)}$  	 \vspace{3pt} \\ \hline  
			Pure	 ($\theta=0^\circ$) &  356	& 403	& 406 & 436 & 647 & 688	 \\ 
			Mixed ($\theta=12^\circ$)	  &  347	& 390	& 393 & 421 & 629 & 666	 \\   
			Expt.~\cite{PDG20} &  $335\pm 75$ & $407\pm 5$ & --- & --- & --- & $689\pm 5$\\ 
			Lattice~\cite{HPQCD-10, HPQCD-12a,HPQCD-12,HPQCD-14} & $394.7\pm 2.4$ & $405\pm6$& $427^{+6}_{-2}$  &  --- & $667^{+6}_{-2}$ & $649\pm31$ \\
			RQM~\cite{CG90b} & --- & ---  & $410\pm20$& ---  & ---  & --- \\
			Sum rules~\cite{BDKMS13} & $387\pm 7$ & $418\pm 9$ & ---  & --- & --- &  --- \\
			BS~\cite{CKWN04} &  $292\pm 25$ & $459\pm 28$& --- & ---  & --- 	&$496\pm20$\\			
            BS~\cite{BCDGPR18} & 385 & ---  &  519(1) & ---  & 709 &  --- \\ 
			LFQM (CJ)~\cite{CJ09} & 326 	& 360	& 349 & 369 & 507 & 529			\\
			LFQM (CJ2)~\cite{CJLR15} &  353 	& 361	& 389 & 391 & 605 & 611			\\   \hline 
			& $f_{\eta_c(2S)}$	& $f_{\Psi'(2S)}$  & $f_{B_c(2S)}$		& $f_{B_c^*(2S)}$	 & $f_{\eta_b(2S)}$  & $f_{\Upsilon(2S)}$ \\ \hline  
			Pure	 ($\theta=0^\circ$) &  318	& 420	& 407 & 477 & 671 & 771	 \\ 
			Mixed ($\theta=12^\circ$)	  &  214	& 274	& 268 & 308 & 443 & 498	 \\   
			Expt.~\cite{PDG20}	  &  ---  & $294(5)$   & --- & --- &---  & $497(5)$\\ 
			Lattice~\cite{HPQCD-14}& --- & --- & --- & --- & --- & $481(39)$ \\
			BLFQ~\cite{LMV17}\footnote{The numbers are read from Fig.~7 of Ref.~\cite{LMV17}.} & 298 & 312 & --- & --- & 525 & 520 \\
			LFD~\cite{LLCLV21} & ---  & $288(6)$ & ---  & ---  & ----  & ---  \\
		\end{tabular}
		\renewcommand{\arraystretch}{1}
	\end{ruledtabular}
\end{table*}

Displayed in the third panel of Fig.~\ref{fig:decay-constant} are our results on the ratio $R_f= f_{2S}/f_{1S}$ obtained 
with $\theta=0^\circ$, $6^\circ$, and $(12\pm 1)^\circ$ cases, 
which are compared with the available experimental data~\cite{PDG20} and the lattice simulations~\cite{HPQCD-14}. 
In particular, we present the results with $\theta = (12\pm 1)^\circ$ as a band to check the sensitivity of the ratio $R_f$ 
on the variation of the mixing angle around $\theta=12^\circ$.
As one can see from the experimental data for $(J/\psi, \Upsilon)$, $R_f$ should be less than unity for heavy meson systems.
The same constraint, $R_f <1$, for light mesons was also discussed in Ref.~\cite{AJ99}.
In our case with the pure state ($\theta=0^\circ$), most heavy mesons except ($D, D_s, \eta_c$) mesons
violate the constraint $R_f <1$.
For the critical mixing ($\theta=\theta_c=6^\circ$) case, one can see that the constraint $R_f<1$ is satisfied for all heavy mesons.
However, the predictions with this critical mixing angle are still not comparable with the available experimental data.
For the mixing angles~$\theta=(12\pm 1)^\circ$, our results $R_f$ for $(J/\psi, \Upsilon)$ are now quite close to the data.
Further experimental measurements on the $2S$ decay constants are, therefore, highly desirable for testing our mixing angle effects.

\begin{figure*}[t]
	\centering
    \includegraphics[width=0.95\textwidth]{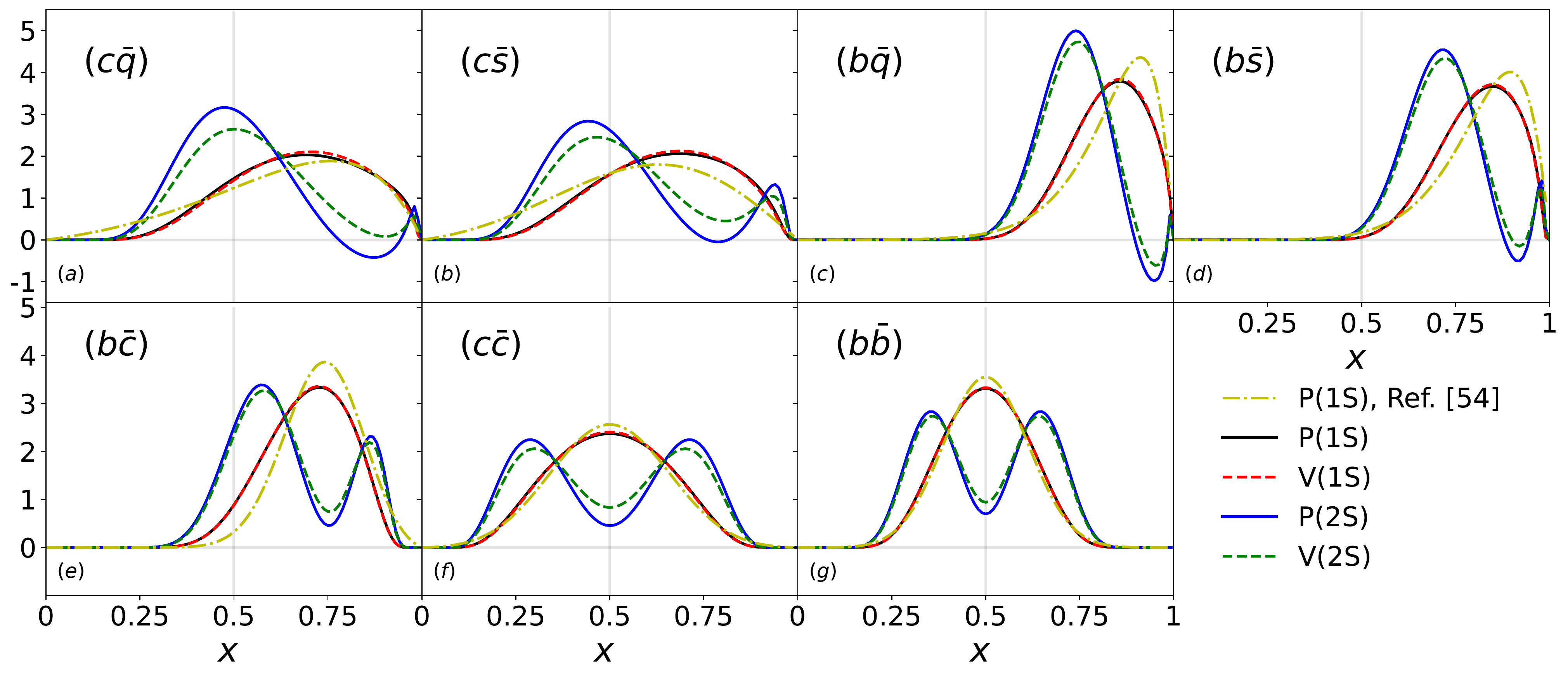}
	\caption{\label{fig:da_mix} 
	Distribution amplitude of the pseudoscalar (P) and vector (V) mesons of 1S and 2S states with the mixing angle $\theta=12^\circ$. 
	The distribution amplitudes of the $1S$ pseudoscalar heavy meson predicted in Ref.~\cite{SDCER20} are
	shown for comparison.}
\end{figure*}

\subsection{Twist-2 distribution amplitudes}
\label{numerical-DA}

Shown in Fig.~\ref{fig:da_mix} are the normalized twist-2 DAs, ${\tilde\phi}^{\rm tw-2}_{P(V)}(x,\mu)$, for the $1S$ and $2S$ state 
heavy pseudoscalar and vector mesons with $\theta=12^\circ$. 
Since the qualitative behaviors obtained with the pure ($\theta=0^\circ$) states are not much different from the mixed case, 
we do not give the results for the pure case in Fig.~\ref{fig:da_mix}.
In this figure, the convention is chosen so that the heavier quark in $(Q \bar{q})$ configuration carries the longitudinal 
momentum fraction $x$ and the lighter quark carries the fraction of $1-x$.
The DAs for the $1S$ state heavy pseudoscalar and vector mesons are given by the black solid and red dashed lines,
respectively.
We also compare our results for $1S$ pseudoscalar heavy mesons with the results of Ref.~\cite{SDCER20} (brown dot-dashed lines)
obtained by employing a continuum approach 
to the hadron bound-state problem.

The DAs for $1S$ state heavy pseudoscalar mesons are not much different from those for the corresponding vector mesons within
our LFQM mainly because the $\beta$ parameters are same for both pseudoscalar and vector mesons.
Although some quantitative differences can be found, in particular, for the $B_c$ and heavy-light mesons, the qualitative behaviors of 
our results for pseudoscalar mesons are similar to those of Ref.~\cite{SDCER20}.
For the $1S$ state heavy quarkonia ($c{\bar c}$, $b{\bar b}$), although both DAs are symmetric under $x\to 1-x$, the shape of the
DA is narrower for the bottomonium than the charmonium. 
However, the DAs for heavy-light systems become more asymmetric and more sharply peaked as the mass difference between the two 
constituents grows.
In particular, one can see that the peak of the DA for heavy-heavy system such as the $B_c$ is more attracted to the center compared to
the heavy-light system.

\begin{figure}[t]
	\centering
	\includegraphics[width=0.7\columnwidth]{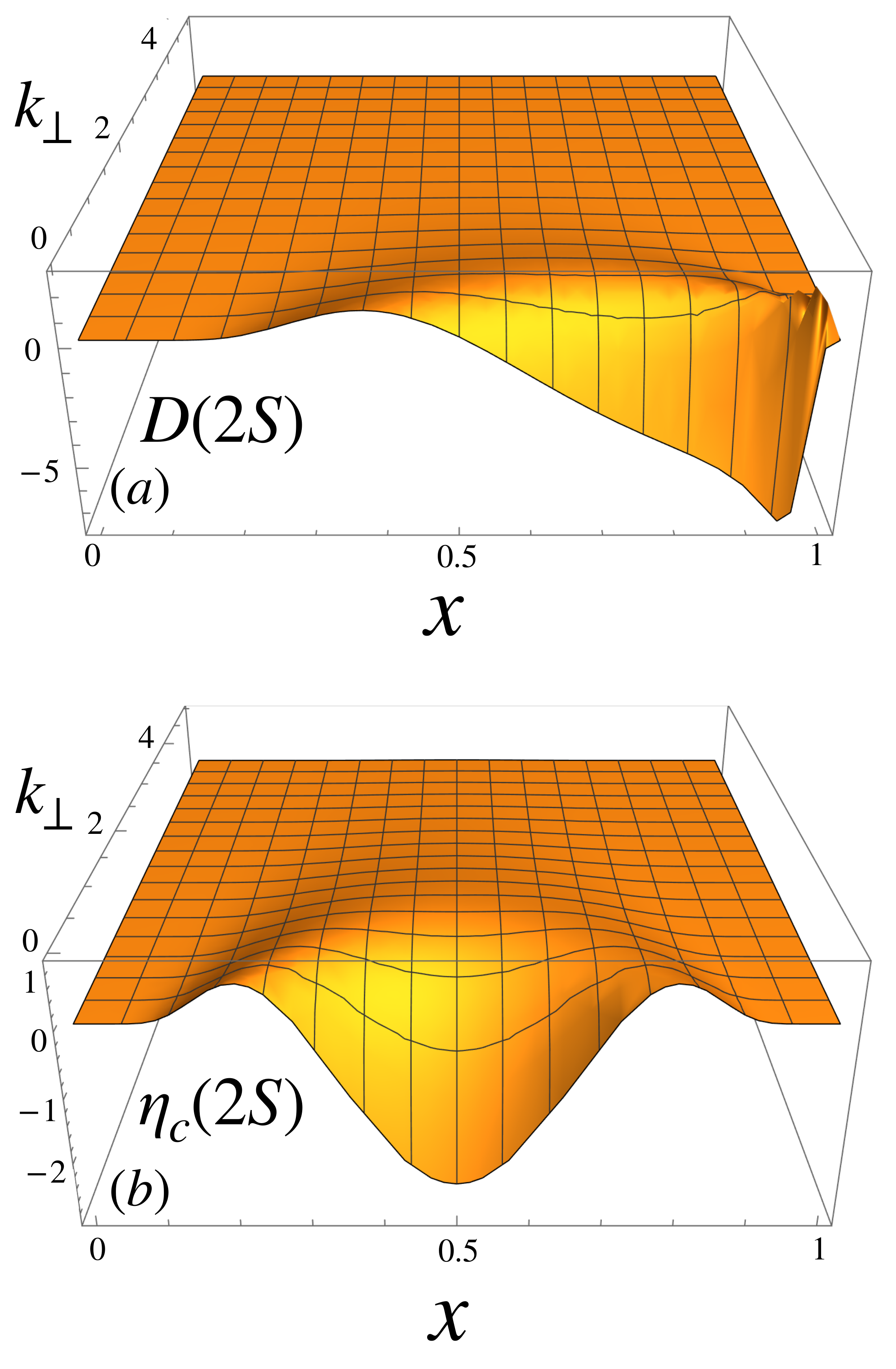}
	\caption{\label{fig:3d_da} 
	3D plots of (a) $\Psi^{\rm tw-2}_{D}(x,{\bf k}_\perp)$ for the $2S$ states of the $D$ meson 
	and (b) $\Psi^{\rm tw-2}_{\eta_c}(x,{\bf k}_\perp)$ for the $2S$ states of the $\eta_c$ meson. }
\end{figure}

The DAs of the $2S$ state pseudoscalar and vector mesons are shown by the blue solid and green dashed lines, respectively, 
in Fig.~\ref{fig:da_mix}.
This shows that the differences between pseudoscalar and vector mesons are more pronounced for the $2S$ states than for 
the $1S$ states. 
This tendency is opposite to that of BLFQ results~\cite{TLMV19} where the differences are more pronounced for the $1S$ states.
We also found that the locations of the two extrema for the quarkonia systems move towards the end points as the quark mass decreases. 
In addition, the valley of the DAs is found to be much lower than those in Ref.~\cite{TLMV19}.
Finally, we mention that the DAs for $2S$ state light meson sector such as $(\pi, K)$ reported in Ref.~\cite{LCGRSZ16} show 
similar qualitative behaviors found in the present work for the $2S$ state heavy mesons.

\begin{table*}[t]
	\begin{ruledtabular}
		\renewcommand{\arraystretch}{1}
		\caption{ \label{tab:xi_moment}
		The $\xi$-moment up to $n=6$ for the $1S$ and $2S$ state heavy pseudoscalar and vector mesons. }
		\begin{tabular}{ccccccccccccccc}
			$(1S)$			&$D$ & $D^*$ & $D_s$ &$D_s^*$   & $\eta_c$ & $J/\psi$ & $B$ & $B^*$ & $B_s$ & $B_s^*$ & $B_c$ & $B_c^*$ & $\eta_b$ & $\Upsilon$ \\ \hline
			$\expval{\xi^1}$ & 0.337 & 0.344 & 0.294 & 0.296 & --- & ---  & 0.644 & 0.646  & 0.614 & 0.614  & 0.390 & 0.390 & --- &--- \\
			$\expval{\xi^2}$ & 0.226 & 0.226 & 0.197 &  0.194  & 0.088 & 0.086  & 0.453 & 0.454 & 0.417 & 0.417  & 0.201 &  0.201 & 0.049 &0.049 \\
			$\expval{\xi^3}$ & 0.145 & 0.144 & 0.114 & 0.112  & --- & ---     & 0.337 & 0.338  & 0.302 & 0.301  &  0.113 &  0.112 & --- &--- \\
			$\expval{\xi^4}$ & 0.108 & 0.107  & 0.083 & 0.080  & 0.018 & 0.017  & 0.261 & 0.262 & 0.228 &0.227 & 0.068& 0.068  & 0.006 &0.006 \\
			$\expval{\xi^5}$ & 0.082  & 0.080  & 0.058 & 0.056 & --- & ---  & 0.209 & 0.210 & 0.178 & 0.177 & 0.043 & 0.043  & --- &--- \\
			$\expval{\xi^6}$ &  0.065 &  0.063 &  0.044 &0.042  & 0.005 & 0.005  & 0.172 & 0.172 & 0.143 &  0.142 &0.029  & 0.028 & 0.001 &0.001 \\ \hline 
			$(2S)$		&$D$ & $D^*$ & $D_s$ &$D_s^*$   & $\eta_c$ & $J/\psi$ & $B$ & $B^*$ & $B_s$ & $B_s^*$ & $B_c$ & $B_c^*$ & $\eta_b$ & $\Upsilon$ \\ \hline
			$\expval{\xi^1}$ &$-0.042$ & 0.071  & 0.015&  0.085 & --- & ---  & 0.426 & 0.452  & 0.411 & 0.433 & 0.275 & 0.285  & --- &--- \\
			$\expval{\xi^2}$ & 0.052 & 0.096  &  0.132&   0.140 & 0.179 & 0.160  & 0.198 & 0.227  & 0.202& 0.224  & 0.160 &  0.162 & 0.099 &0.094 \\
			$\expval{\xi^3}$ & $-0.004$& 0.034 & 0.055 & 0.064  & --- & ---  & 0.094 & 0.121 & 0.112 &  0.130& 0.101 &  0.100& --- &--- \\
			$\expval{\xi^4}$ & 0.012 & 0.033  & 0.065 &0.064  & 0.048 & 0.042  & 0.043 & 0.067 & 0.070 & 0.084 & 0.071& 0.069 & 0.016 &0.015 \\
			$\expval{\xi^5}$ & 0.006 & 0.022 & 0.047 & 0.045  & --- & ---  & 0.017&  0.037 &  0.048 &0.059  & 0.051& 0.049  & --- &--- \\
			$\expval{\xi^6}$ & 0.010 & 0.020 & 0.044 &  0.040& 0.016 & 0.014  & 0.004 &  0.021 & 0.036 &  0.045& 0.038& 0.036  & 0.003 &0.003 \\
		\end{tabular}
		\renewcommand{\arraystretch}{1}
	\end{ruledtabular}
\end{table*}

The normalized twist-2 pseudoscalar and vector meson DAs are rewritten as
\begin{equation} \label{eq:27}
{\tilde\phi}^{\rm tw-2}_{P(V)}(x,\mu) = \int_0^{|{\bf k}_\perp|<\mu} d^2 {\bf k}_\perp \Psi^{\rm tw-2}_{P(V)}(x,{\bf k}_\perp),
\end{equation}
where the LF wave function corresponding
to ${\tilde\phi}^{\rm tw-2}_{P(V)}(x,\mu)$ is denoted as
$\Psi^{\rm tw-2}_{P(V)}(x,{\bf k}_\perp)$.
Shown in Fig.~\ref{fig:3d_da} are the 3-dimensional (3D) plots of $\Psi^{\rm tw-2}_{D}(x,{\bf k}_\perp)$ and 
$\Psi^{\rm tw-2}_{\eta_c}(x,{\bf k}_\perp)$ for the $2S$ states of $D$ and $\eta_c$ mesons, respectively. 
Equation~(\ref{eq:27}) implies that the normalized twist-2 DAs 
${\tilde\phi}^{\rm tw-2}_{P(V)}(x,\mu)$ 
shown in Fig.~\ref{fig:da_mix}
are obtained by the ${\bf k}_\perp$-integration of $\Psi^{\rm tw-2}_{P(V)}(x,{\bf k}_\perp)$.
In our LFQM calculation with the Gaussian wave functions, we observe that $|{\bf k}_\perp|\to\infty$ corresponds to
the ultra violet (UV) cutoffs ($k^{\rm max}_\perp$) or energy scale $\mu$ around 2~GeV for ($D_{(s)}$, $B_{(s)}$, $\eta_c$), 3~GeV for 
$B_c$, and 4~GeV for $\eta_b$, respectively. 
In other words, the wave functions for heavy-heavy systems have longer transverse momentum tails than for the 
heavy-light systems. 
On the other hand, the wave functions for heavy-light systems have deeper negative valleys than for heavy-heavy systems.
This property in the twist-2 light-front wave function explains why only the twist-2 DAs for the $2S$ state heavy-light system
have negative regions. 
Of course, the small UV cutoff such as $|k^{\rm max}_\perp|<2$ GeV for heavy-heavy system
may cause ${\tilde\phi}^{\rm tw-2}_{P}(x)$ to have negative regions as well. 
Similar observations are made for the $2S$ state vector meson DAs.

In Table~\ref{tab:xi_moment}, we provide the $\xi$-moments defined in Eq.~(\ref{eq:xi}) up to $n=6$ for both $1S$ and $2S$
state heavy mesons. 
For heavy quarkonia, the odd-$n$ moments vanish because of the symmetric shape in their DAs. 
For heavy-light mesons, the odd-$n$ moments reflect the asymmetry of the DAs coming from the mass difference between 
the quark and antiquark. 
As one can see, the first $\xi$ moment decreases as $(m_Q - m_{\bar q})$ gets smaller.
For example, we have $\expval{\xi^1}_{B_q(1S)}=0.644$, $\expval{\xi^1}_{B_s(1S)}=0.614,$ and $\expval{\xi^1}_{B_c(1S)}=0.390$.
The DAs for $1S$ state mesons can be well reproduced with the first few $\xi$ moments up to $n=4$.
However, those for $2S$ state mesons require much higher $\xi$ moments beyond $n=6$ to reproduce the full results.
For the $2S$ state $D$ mesons, the first and third moments $\expval{\xi^n}_{D(2S)}$ $(n=1,3)$ have negative values. 
This originates from the fact that the DA of $D(2S)$ occupies more in the $0<x<0.5$ domain compared to the other DAs. 
In particular, our predictions of $\left(\expval{\xi^2},\expval{\xi^4},\expval{\xi^6} \right)=(0.179,0.048,0.019)$ for the $2S$ state of $\eta_c$
are comparable to $(0.16, 0.046, 0.016)$ of Ref.~\cite{Braguta07}, which are obtained from the leading twist light-front wave functions 
with the Cornell potential.

\subsection{Electromagnetic form factors and radii}

\begin{figure*}[t]
	\centering
	\includegraphics[width=0.95\textwidth]{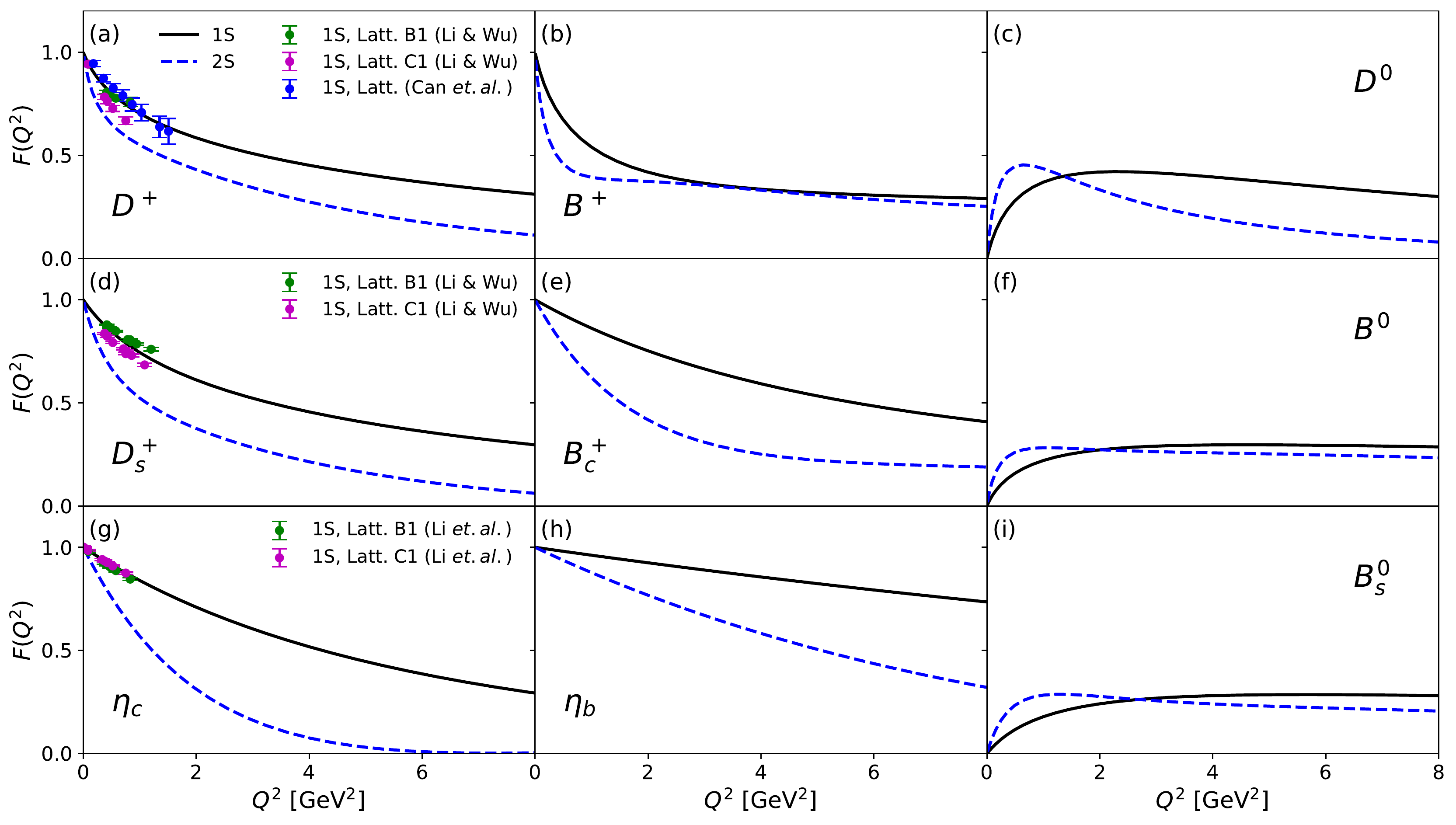}
	\caption{\label{fig:form_factor1} 
	Electromagnetic form factor of the $1S$ and $2S$ state heavy pseudoscalar mesons.
	The available lattice simulations are also shown by blue circles for the results of Ref.~~\cite{CEOOT12}, and 
	green circles and magenta circles for B1 and C1 ensembles of Refs.~\cite{LW17,LLW20}, respectively. 
	Only one quark contribution is considered for quarkonia.}
\end{figure*}

In Fig.~\ref{fig:form_factor1}, we present the electromagnetic form factors of $1S$ (solid lines) and $2S$ (dashed lines) state heavy pseudoscalar mesons obtained with $\theta=12^\circ$.
For comparison, we show the available lattice simulation data of Refs.~\cite{CEOOT12,LW17,LLW20}.
Since the form factors of heavy quarkonia ($\eta_c, \eta_b$) obtained from both quark and antiquark contributions
vanish, we show only the contribution from the quark part 
for the comparison with the available lattice simulation results.
This shows that our results for the $1S$ state $(D^+, D^+_s, \eta_c)$ mesons match well with the lattice simulation results.
We find that the form factor of $2S$ state mesons are in general steeper than those for the corresponding $1S$ state mesons.
In a heavy-light system such as $(D^+_{(s)}, B^+)$, the main contribution to the form factor for the region of $Q^2 > 6$~GeV$^2$ 
comes from the heavy quark and the light quark contribution is negligible at high $Q^2$ regions. 
On the other hand, for the form factor of the $B_c$ meson, both $b$ and $c$ quark contributions are almost equally important for the intermediate $Q^2$ region.

\begin{figure*}[t]
	\centering
	\includegraphics[width=0.95\textwidth]{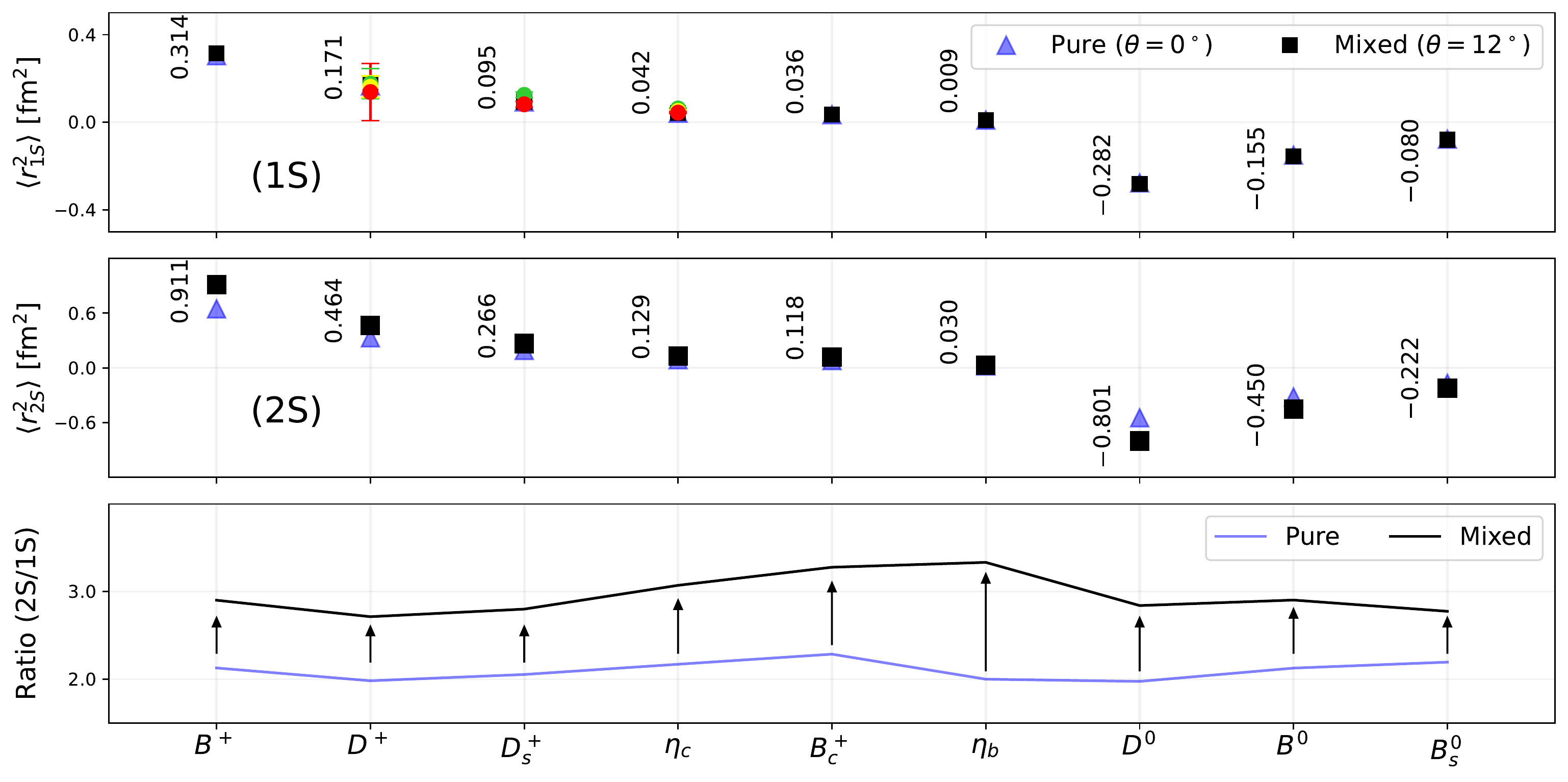}
	\caption{\label{fig:radius} 
	(Upper and middle panels) Charge radii of the $1S$ and $2S$ state heavy mesons with various quark flavor contents. 
	Our predictions for $1S$ heavy mesons have a good agreement with the lattice simulation data of Refs.~\cite{CEOOT12,LW17,LLW20,DER06}. 
	(Lower panel) The ratio of the charge radii $R_r= \Braket{r^2_{2S} }/ \Braket{r^2_{1S} }$ for various heavy mesons. 
	The charge radii in the mixed state are larger than the pure ones. }
\end{figure*}

\begin{table*}[t]
	\begin{ruledtabular}
		\renewcommand{\arraystretch}{1}
		\caption{\label{tab:radii}
		Charge radii of the $1S$ and $2S$ pseudoscalar heavy mesons in the units of fm$^2$. }
		\begin{tabular}{lccccccccc}
			$\Braket{r^2}$	& $B^+(1S)$ 	& $D^+(1S)$ & $D_s^+(1S)$ & $\eta_c(1S)$  &  $ B_c^+(1S)$  &  $\eta_b(1S)$ &  $D^0(1S)$ & $B^0(1S)$  &  $B_s^0(1S)$ \\ \hline
			Pure	 ($\theta=0^\circ$)	 	& 0.304	 & 0.166	& 0.093& 0.041 & 0.035	& 0.010	& $-0.277$ 	& $-0.150$  & $-0.077$\\  
			Mixed ($\theta=12^\circ$)		   & 0.314	 & 0.171	& 0.095	& 0.042 & 0.036	& 0.009	& $-0.282$ 	& $-0.155$  & $-0.080$\\  
			Lattice, Linear fit~\cite{CEOOT12}			 & --- & $0.138(13)$ &--- &--- & --- &--- &--- & --- & --- \\
			Lattice, Quadratic fit~\cite{CEOOT12}			 &  --- & $0.152(26)$ &--- &--- & --- &--- &--- & --- & --- \\
			Lattice, B1~\cite{LW17,LLW20}	  & --- & $0.162(49)$  & $0.082(13)$& $0.052(4)$& --- & --- & --- & --- & --- \\
			Lattice, C1~\cite{LW17,LLW20}	  & --- & $0.176(69)$ & $0.125(13)$& $0.044(4)$ & --- & --- & --- & --- & --- \\
			Lattice,~\cite{DER06} &	---& ---& ---& 0.063(1) & --- & --- & --- & --- & --- \\
			CCQM~\cite{MDTF21} & --- & 0.255 & 0.142 & --- & ---& --- & --- & --- & --- \\
			LFQM~\cite{Hwang01} 	& 0.378 & 0.184	& 0.124	& --- & 0.0433 & --- &$-0.304$	&   $-0.187$ & $-0.119$ \\ 
															& 0.496  & 0.248	& 0.181 &  --- & ---	& ---& $-0.496$	& $-0.248$ & $-0.181$  \\ 
			BLFQ~\cite{LMZV15} 	& 	---	& ---	& --- & 0.038(5) & --- & 0.0146(8)& --- &   --- &  --- \\		\hline 
			$\Braket{r^2}$	& $B^+(2S)$ 	& $D^+(2S)$ & $D_s^+(2S)$ & $\eta_c(2S)$  &  $ B_c^+(2S)$  &  $\eta_b(2S)$ &  $D^0(2S)$ & $B^0(2S)$  &  $B_s^0(2S)$ \\ \hline
			Pure	($\theta=0^\circ$)   	& 0.647	 & 0.329	& 0.191 & 0.089 & 0.080	& 0.020	& $-0.547$ 	& $-0.319$  & $-0.169$\\  
			Mixed ($\theta=12^\circ$)	   & 0.911	 & 0.464	& 0.266	& 0.129 & 0.118 & 0.030	& $-0.801$ 	& $-0.450$  & $-0.222$\\   
			BLFQ~\cite{LMZV15} 	& ---	& ---	& --- & 0.1488(5) & --- & 0.0510(8)& --- &   --- & --- \\
		\end{tabular}
		\renewcommand{\arraystretch}{1}
	\end{ruledtabular}
\end{table*}

Figure~\ref{fig:radius} presents the calculated charge radii $\Braket{ r^2_{nS} }$ of the $(1S, 2S)$ state heavy mesons obtained with $\theta=0^\circ$ and 
$\theta=12^\circ$ cases by triangles and boxes, respectively.
Since there is no experimental data, our results are compared with the lattice simulation results of Refs.~\cite{CEOOT12,LW17,LLW20,DER06}.
The full results are summarized in Table~\ref{tab:radii} with other model predictions from Refs.~\cite{MDTF21,LMZV15,Hwang01}. 
We find that our predictions are overall in a good agreement with the lattice results.
Since the heavy quark is sitting near the center of a meson while the light quark is moving actively, the radius of a heavy-light meson would be
mostly governed by the motion of the light quark.
This can be noticed by comparison with the results of the LFQM of Ref.~\cite{Hwang01}.

We find that the charge radii for $2S$ state mesons are more sensitive to the mixing angle than for the $1S$ state cases.
As shown in Fig.~\ref{fig:radius}, the charge radii for $2S$ state mesons are larger in the mixed case than those in the pure one.
The larger radii can be understood from the $2S$ wave functions given in Fig.~\ref{fig:wavefunction}.
Namely, the reduction of the wave function at the origin results in a larger radius since the wave function is more spreading to a larger distance.
This is the opposite behavior of the decay constants shown in Fig.~\ref{fig:decay-constant}, which is proportional to the wave function at the origin.
Therefore, these observables reflect the structure of the wave functions from different points of view.

\section{Summary} \label{sec:summary}

In the present work, we have investigated $1S$ and $2S$ state heavy mesons employing the pure and mixed harmonic oscillator wave functions. 
We invoked the variational principle adopting the linear plus Coulomb potential, and treated the hyperfine interaction perturbatively as a contact 
term to distinguish vector and pseudoscalar mesons. 
The variational principle allowed us to obtain a constraint for model parameters. 
With the fixed quark masses, all model parameters are determined by two meson masses and this led us to predict and test other physical quantities.

We have analyzed the mass spectra, decay constants, distribution amplitudes, electromagnetic form factors, and charge radii of 
the $1S$ and $2S$ state heavy mesons. 
As for the mass spectra, our predictions are in a good agreement with the available experimental data. 
Although no apparent difference is found for the masses of the $1S$ heavy mesons in the pure and mixed cases, 
the predicted masses of the $2S$ heavy mesons are appreciably modified and have a better agreement with the available data 
when the mixing is introduced.
Our results support the speculation that the newly observed $D_s(2590)$ by the LHCb Collaboration~\cite{LHCb-20d} can be interpreted 
as the radially excited $D_s(2S)$ state.

We also observe that the mass gaps between the $1S$ and $2S$ state mesons are around 600~MeV and they are not sensitive to 
the flavor contents of mesons.
In LFQM, we found that mass gaps of pseudoscalar mesons can be made larger than those of vector mesons regardless of
the quark flavor contents only if we use the mixing angle $\theta \geq \theta_c=6^\circ$.
Such behavior can be explained by Eq.~(\ref{eq:mass_gap}), which shows that the hierarchy appears in the opposite direction without mixing.

The mixing effects are crucial to understand the properties of $2S$ state mesons.
As for the decay constants, we could obtain a good agreement with the experimental and lattice simulation data for $1S$ state mesons
even without the mixing effects.
However, for the 2S states, the mixing effects are essential to get the correct order of decay constants.
By introducing a small mixing, we noticed that the ratio $R_{\Phi}=\Phi_{2S}^2/\Phi_{1S}^2$ becomes smaller than unity. 
The optimum value of the mixing angle is obtained as $12^\circ$ to cover both the charm and bottom flavors of the heavy quark.

For the DAs of the $1S$ states, our prediction is found to be similar to those reported in Ref.~\cite{BCDGPR18}.
For the 2S states, some DAs have the nodal structure arising from the structure of the wave functions. 
We note that the difference between DAs for vector and pseudoscalar mesons are more pronounced for the $2S$ states.
In addition, we find that the DAs are saturated up to several GeV for the transverse momentum and it has longer tails for 
mesons with heavier quarks.
For completeness, the corresponding $\xi$-moments up to $n=6$ are computed in the present work.

The electromagnetic form factors and charge radii for $D$, $D_s$, and $\eta_c$ mesons are also computed and found to be comparable 
with the available lattice simulation data of Refs.~\cite{CEOOT12,LW17,LLW20}.
The mixing effects lead to larger radii of the $2S$ states since their wave functions are more spread in space, which results 
in the reduction of the wave functions near the origin.
This is opposite to the behavior of the decay constants that are reduced by the mixing.

In this work, we have focused on the heavy meson sector in LFQM.
However, a combined analysis for both light and heavy meson sectors is also of great importance for scrutinizing the dependence of 
physical quantities on quark masses.
In the future work, we would consider smearing the hyperfine interaction to treat it nonperturbatively as a part of the entire Hamiltonian for 
the application of the variational principle.
A global analysis would be also required for more rigorous investigations to discuss the uncertainties of the model parameters. 
While the investigation along this directions is under progress, more precise measurements on the physical properties of heavy mesons 
as well as observations of undiscovered heavy meson states are essential to test phenomenological models on the structure of heavy mesons.

\acknowledgements

We are grateful to Yongwoo Choi for helpful discussions at the early stage of this work.
A.J.A. was supported by the Young Scientist Training (YST) Program at the Asia Pacific Center for Theoretical Physics (APCTP) through the Science and Technology Promotion Fund 
and Lottery Fund of the Korean Government and also by the Korean Local Governments -- Gyeongsangbuk-do Province and Pohang City.
The work of H.-M.C. was supported by the National Research Foundation of Korea (NRF) under Grant No. NRF- 2020R1F1A1067990.
The work of C.-R.J. was supported in part by the U.S. Department of Energy (Grant No. DE-FG02-03ER41260). 
The National Energy Research Scientific Computing Center (NERSC) supported by the Office of Science of the U.S. Department of Energy 
under Contract No. DE-AC02-05CH11231 is also acknowledged. 
Y.O. was supported by NRF under Grants No. NRF-2020R1A2C1007597 and No. NRF-2018R1A6A1A06024970 (Basic Science Research Program).
The hospitality of the APCTP Senior Advisory Group is gratefully acknowledged.

\end{document}